\documentclass[showkeys,preprintnumbers]{revtex4}

\usepackage{graphicx}
\usepackage{amssymb}
\usepackage{amsmath}
\usepackage{multirow}
\usepackage[utf8]{inputenc}
\usepackage[english]{babel}
\usepackage{amsthm}
\newtheorem{pp}{Proposition}[section]
\newtheorem{lm}[pp]{Lemma}
\newtheorem{cl}[pp]{Corollary}
\newtheorem{df}[pp]{Definition}
\begin{document}
\newcommand{\dfn}[1]{\begin{df}#1\end{df}}
\newcommand{\prp}[1]{\begin{pp}#1\end{pp}}
\newcommand{\lem}[1]{\begin{lm}#1\end{lm}}
\newcommand{\crl}[1]{\begin{cl}#1\end{cl}}
\newcommand{\prf}[1]{\begin{proof}#1\end{proof}}
\newcommand{\q}[1]{``#1''}
\newcommand{\lrbr}[1]{\left\lbrace#1\right\rbrace}
\newcommand{\lrp}[1]{\left( #1\right)}
\newcommand{\lsmatr}[1]{\left\lbrace\begin{smallmatrix}#1\end{smallmatrix}\right.}
\title{A General Theory of Concept Lattice (I):\\ 
\small{Emergence of General Concept Lattice}}
\author{Tsong-Ming Liaw}
\email{ltming@gate.sinica.edu.tw} 
\affiliation{Institute of Physics, Academia Sinica, Taipei, Taiwan 11529}
\author{Simon C. Lin}
\email{Simon.Lin@twgrid.org, Simon.Lin@cern.ch}
\affiliation{Institute of Physics and ASGC, Academia Sinica,  Taipei, Taiwan 11529}
\begin{abstract}
As the first part of
the treatise on {\bf A General Theory of Concept Lattice (I$\sim$V)},
this work develops the {\it general concept lattice}
for the problem concerning categorisation of 
objects according to their properties.
Unlike the conventional 
approaches, such as the formal concept lattice 
and the rough set lattice,
the general concept lattice is designed to
adhere to the general principle that 
the information content 
should be invariant regardless
how the variables/parameters are presented.
Here, one will explicitly
demonstrate the existence of such a construction 
by a sequence of fulfilments
compatible with the conventional
lattice structure.
The general concept lattice 
promises to be a comprehensive categorisation
for all the distinctive object classes according to 
whatever properties they are equipped with.
It will be shown that one can always regain 
the formal concept lattice and rough set lattice
from the general concept lattice. 
\end{abstract}
\maketitle
\section{introduction}\label{one}
It is commonly accepted that the formal concept analysis ({\bf FCA}) \cite{Wi82,GW99,Wi05} and the rough set theory ({\bf RST}) \cite{Pa82,Pa91} are important approaches to the Big Data Analytics.
One can superficially tell such importance
from the rapid growth of interest in both fields, 
e.g., the inclusion of the {\bf FCA} and/or {\bf RST} topics
 in many international conferences and workshops.
Although {\bf FCA} and {\bf RST} might have been motivated differently, it is believed that they are mutually 
expressible \cite{Ke96,DGO01,GD02,DG03,YY04,Wa05}. 
Historically, the {\it formal concept lattice} (FCL) is a native part for {\bf FCA} since {\bf FCA} {\it had its origin in activities of restructuring mathematics, in particular, mathematics order and lattice theory} \cite{Wi05}. On the other hand, the {\it rough set lattice} (RSL) has also been accomplished \cite{DG03,YY04} in terms  of the {\it modal logic} operators which cope with {\bf RST} in the binary version.
Although the two concept lattices 
may be seen as dealing with different categorisations,
one may articulate them of the same object collection $G$ 
and the same attribute set $M$
related in a {\it unique} information system, 
i.e. the same formal context. 
However,
the systematic incorporation of 
the FCL and the RSL in the problem analysis 
remains unobserved.

The treatise 
\q{A General Theory of Concept Lattice (I$\sim$V)} 
is initiated by  
the general idea that 
the information content 
should be invariant 
regardless how the variables/parameters are presented.
Notice the fact that the conventional employment of objects and 
attributes are intuitively {\it different}.
Unlike the objects, which are individuals, the 
attributes as the object property description 
can overlap per {\it conjunction}.
Thus, for general characterisation of logic problems
it is essential also to take into accout
the {\it composite} attributes based on
the operations $\lrbr{\mbox{conjunction, disjunction, negation}}$. 
There is no reason to stress
the priority of the {\it simple}  
attributes in $M$ over the composite attributes  
since one may always rename 
the composite attributes into new {\it simple} (in contrast to composite) ones, which may
render those originally {\it simple} attributes {\it composite}.
Instructed by such a principle, 
the general concept lattice ({\bf GCL})  
thus speaks of the {\it generalised attribute}s, which 
incorporate both the simple and the composite attributes 
constructable out of the members of $M$.

In this part (I) 
the existence of {\bf GCL} subject to the formal context will be
demonstrated, giving rise to
a more informative structure than 
the conventional RSL and FCL.
The {\bf GCL} provides a comprehensive categorisation 
for whatever objects discernible by the formal context 
since in its construction 
one can exhaust all the generalised attributes, 
as will be clarified in the part (II) \cite{LLJD12-2}.
By contrast, the FCL is devised for those
attributes given as conjunctions of the members of $M$
and the RSL for those disjunctions.
Remarkably, the comprehensive categorisation in the {\bf GCL} 
then brings about a simple framework for 
all the implications extractable from the lattice structure.
While these implications are considered between two generalised attributes, 
those implications deducible from the FCL and RSL can be regained when  
one restricts oneself to the attributes of conjunction type and disjunction type, respectively.
The simple framework for all the implications  
enabled by the {\bf GCL}
in fact inspires the idea to represent 
any implication relation 
by one single attribute, thereby rendering 
the primary deduction system ({\bf PDS}), see 
the part (III) \cite{LLJD12-3},
which is a simplified systems where 
logic statements only take into account the properties 
on a single subject by means of predicates.
The {\bf PDS} is {\it algebraically manipulable} 
in the sense that the deductions are solely 
Boolean algebras bypassing any of axioms.
Note that being able to achieve an algebraically manipulable deduction 
is rather practical.
One may then manage to resolve any logic problem mindlessly
if it could be arranged according to the syntax of
the algebraically manipulable deduction.
Indeed, certain prevalent puzzles  
will be shown to be resolved in the {\bf PDS} in this manner.
Another point is that tautology in effect occurs
whenever the algebraically manipulable deduction 
results in the Truth {\bf 1}, which is essentially what all the axioms should end up. 
It can be shown that the {\bf PDS} adheres 
to the classical logic rules since the Hilbert axioms 
all turns out to be tautology when restricted to the {\bf PDS}.

In the parts (IV) and (V) of this treatise
\cite{LLJD12-4,LLJD12-5}, 
efforts are devoted to render
the algebraically manipulable deduction 
more realistic such that it can  
cope with the conventional reasoning. The following
two points are of concern.
\begin{itemize}
\item  
Dealing with pure attribute-typed logic statements in the {\bf PDS}
is na{\"i}ve since 
in reality every assertion, where a property assignment  
has to be ascribed to a {\it definite} referential 
attribute set based on which 
a collection of judgements altogether forms the statement. 
\item  The {\bf PDS} as concerned with 
the deduction on one sole subject (object class) 
cannot be sufficiently expressive in formalising a 
general logic statement; one is looking forward to 
the further extensions. 
\end{itemize} 
For the first point,  
various attribute-type logic statements,
though ascribed to different referential attribute sets, 
may become the same one.
In fact, such a problem reminds one about the conventional indifference of the set unions 
$X^c\cup X=Y^c\cup Y=(X^c\cap Y^c)\cup (X\cap Y^c)\cup (X^c\cap Y)\cup (X\cap Y)$ and so forth.
Therefore, one argues \cite{LLJD12-4} that
the notion of {\it finite resolution} 
enters the syntax of {\bf PDS} and participates in the deduction as a novelty.
The {\bf PDS} incorporating the {finite resolution} 
will then be referred to as  
the {\it resolvable} {\bf PDS}, which however exhibits 
anomalies for some of the Hilbert Axioms, entailing that
the classical logic comprises certain counter-intuitiveness.
As for the second point, 
the secondary deduction system ({\bf 2DS}) \cite{LLJD12-5}
is designed to take into account those logic statements 
relating predicate forms among different subjects.
Note that, despite involving the {\bf PDS} syntax as its
particular statements,
the {\bf 2DS} can be developed in parallel to the {\bf PDS}.
Thus, one does not anticipate different deduction rules from 
the {\bf 2DS} than from the {\bf PDS}. 
Moreover, the resolvable {\bf 2DS} can likewise arise 
from the reasoning incorporating finite resolution, 
which can be traced back
to the same origin that elicits the resolvable {\bf PDS}.

To start with in this part (I),
the {\bf GCL} is a structure 
ordering lattice nodes marking the {\it general concepts}
by means of Galois connections. 
The general concept
takes the 2-tuple representation 
({\it general extent}, 
 {\it general intent}), ({\it Ge},{\it Gi}) for short,
which is similar to
the formal concepts employed in FCL 
and in RSL. 
Below,
one will demonstrate the existence of such {\bf GCL} through 
the accomplishment of the following goals. 
\begin{description}
\item[G1]
The general extent runs over 
all the distinctive object classes recognised by the formal context. 
The general intent plays the role of properties 
of a definite object class
and can be characterised by a pair of attributes
referred to as
the {\it generalised rough set property} ({\it Grsp}) and the {\it generalised formal concept property} ({\it Gfcp}). 
The 2-tuples ({\it Ge},~{\it Grsp}) and 
({\it Ge},~{\it Gfcp})
in effect satisfy
the generalisation of lattice conditions, respectively, 
employed for the RSL \cite{DG03,YY04} and for the FCL \cite{Wi82,GW99,Wi05}.
\item[G2]
The {\bf GCL} constitutes a Hasse diagram with 
a power set structure, where the nodes in the Hasse diagram can be related by means of
set inclusion relations among the object classes discernible 
by the formal context.
\item[G3]
Both RSL and FCL can be regained from the {\bf GCL}
as particular features.
%
\item[G4]
The {\bf GCL} allows for {\it deterministic construction} based on a {\it particular collection} of general concepts to be described later.
\item[G5]
The {\bf GCL} manifests a conjugate relation between its RSL- and FCL- counterparts and
consequently emerges as a {\it self-dual} lattice.
\end{description}

The formulation of the work at hand 
differs from the traditional approaches (typically, \cite{Wi82,GW99,DG03,YY04}) mainly by the employment of the {\it generalised} attributes. 
For concreteness, let $M$ be the attribute set under consideration
then the {\bf GCL} takes into account {\it generalised attributes}
$M^\ast$ which are all the possible compositions of members from $M$ 
by means of all the basic Boolean operations.
Thus, the information provided by 
the composite attributes {$M^\ast \backslash M$} could be as significant 
as the information given in terms of {\it simple} (in contrast to the composite ones) attributes $M$. 
In Sec.~\ref{two} some preliminaries of the {\bf RST} and {\bf FCA} are revisited, where
one attempts an equal-footing treatment 
for both theories
based on 
their original conventions 
with minimal notation refinements. 
This is to establish 
a convenient framework 
above which one may 
differentiate the {\bf GCL} with the RSL and FCL.
In Sec.~\ref{three}, 
it is shown that both the FCL- and RSL- intents could have been 
represented by {\it single composite attributes} without loss of generality. 
Hence, the consideration to generalise
the attribute domain from $M$ to $M^\ast$ 
in the problem setting becomes intuitive.
Moreover, the sign of the {\bf GCL} is revealed from
some inconsistency observed in 
the conventional FCL and RSL approach.
In order to retain the consistency, 
one inevitably takes into account 
the full correspondence between the objects 
and generalised attribute set $M^\ast$, which then gives rise to
the {\bf GCL}.
It turns out that the {\bf GCL} ends up 
a comprehensive categorisation 
since its object classes range over 
all the possible object sets discernible by the formal context.
In addition, the nodes on the {\bf GCL} can be ordered as 
general concepts.
While each object class is 
identified with a {general extent},
the corresponding {\it general intent}
as its associate property
may be acquired from the generalisation of 
FCL-intent and RSL-intent.
Although all the general intents can 
be directly determined per read out from the formal context, as will be demonstrated in the next paper,
one chooses here a constructive approach for  
them by means of certain {\it irreducibility} conditions.
It is through the design of such irreducibility conditions 
that one can reveal how  
the traditional RSL and FCL being regained as part of {\bf GCL}.
In Sec.~\ref{four}, results are summarised to affirm the 
accomplishments of goals {\bf G1} to {\bf G5}. 
There are also discussions concerning 
further developments for the general theory of concept lattice. 
\section{preliminaries and beneath}\label{two}
The formal context as stated in the {\bf FCA} approach \cite{Wi05} is defined as 
{\it a set structure  $K:=(G,M,I)$, for which $G$ and $M$
are sets while {$I$ is binary relation between 
$G$ and $M$}, i.e.
$I\subseteq G\times M$; the elements 
of $G$ and $M$ are called (formal) objects 
and (formal) attributes, respectively, 
and $gIm$, i.e. $(g, m)\in I$, is read: the object $g$ has the attribute $m$}.
However, for the {\bf RST} theorists there is a different convention, governing similar 
functions, where the {\it information system} ({IS}) $(U,V,R)$ \cite{Pa82} is often employed as a synonym of formal context.  
%
Hence, in order to prevent from confusion caused by simultaneously treating both theories 
there is a compromise that preserves most of the original notations employed in {\bf FCA} and {\bf RST} as follows.
\dfn{A formal context is a set structure $K:=(G,M,R)$ 
 (let it also be denoted by $F(G,M)$),
 for which $G$ and $M$ are sets 
 while $R$ is binary relation between $G$ and $M$, i.e.
 $R\subseteq G\times M$; 
 the elements of $G$ and $M$ are called {\it formal} objects and {\it formal} attributes, respectively, and $gRm$, i.e. $(g, m)\in R$, is read: the object $g$ has the attribute $m$.} 
Here, the notation {$F(G,M)$} is to emphasise that $G$ and $M$ are related, conceived through the formal context $F$, which is instructive whenever more than one formal contexts are simultaneously treated.
In particular, the definition given 
in Ref.~\cite{Wi05} is modified into {\bf Definition~2.1} in the following manner.
\begin{itemize}
\item
The original notation $I$ used in {\bf FCA} is reserved for the {\it derivation operator}, which is typically a map {\it from subset to subset}.
$R$ is a binary operation which relates 
a single object (attribute) to an attribute-set (object-set), 
denoted by
$gRm$ \cite{Wi82,GW99,Wi05} where $g^R$ is an attribute-set and $m^R$ is an object-set.
\begin{itemize}
\item
$m\in g^R$ ($g\in G \mapsto g^R \subseteq M$): 
the object $g$ {\it has the attribute} $m$,
\item
$g\in m^R$ ($m\in M \mapsto m^R \subseteq G$):  
$g$ is one of the objects that {\it possess the attribute} $m$.
\end{itemize}
Based on $R$, one may define all the derivation operators for both {\bf FCA} and {\bf RST}, i.e. $I$, $\Box$ and $\Diamond$, as will be clarified in {\bf Definition~2.2}.
\item
The word {\it formal} is kept for the purpose to distinguish 
objects ($G$) and attributes ($M$),
in contrast to the traditional approaches where only set operations are employed for both objects and attributes.
\begin{itemize}
\item Members of $G$ 
are to be treated {\it formally like objects}.
In practice, there is a collection of individuals 
to be categorized into classes. 
Thus, the operations for objects
are those which are apt to manipulate set relations,
say intersection ($\cap$), union ($\cup$) and complementarity ($\lbrace \rbrace^c$). 
\item Members of $M$ 
are to be treated  {\it formally like attributes}.
It is natural to extend simple attributes into  
the composite attributes by 
means of the Boolean operations.
Here,\\
the product \q{$\cdot$} ($\prod$) is employed for the {\it conjunction}, i.e. the  logical {\bf AND},\\
the summation \q{$+$} ($\sum$) is employed for the {\it disjunction}, i.e. the logical {\bf OR},\\ 
the unary operation \q{$\neg$} is employed for the {\it negation}, i.e. the logical {\bf NOT}. 
\end{itemize}
\end{itemize}
%
A brief summary  
in terms of the above new notation for the traditional approaches in {\bf FCA} and {\bf RST} is in order. \\
\begin{df}[\cite{Wi82,GW99,Wi05,DG03,YY04}]
Given a formal context $F(G,M)$,  
the derivation operators 
are  
\begin{eqnarray}
X\subseteq G &\mapsto& X^I=\lbrace m\in M \mid gRm,\ \forall g \in X \rbrace\subseteq M,\nonumber\\
Y\subseteq M &\mapsto& Y^I=\lbrace g\in G \mid gRm,\ \forall m \in Y \rbrace\subseteq G,
\label{eq1}\\ 
X\subseteq G &\mapsto& X^{\Box}=\lbrace m\in M \mid \forall g \in G, gRm \implies g\in X \rbrace\subseteq M,\nonumber\\
Y\subseteq M &\mapsto& Y^{\Box}=\lbrace g\in G \mid \forall m \in M, gRm \implies m\in M \rbrace\subseteq G,
\label{eq2}\\
X\subseteq G &\mapsto& X^{\Diamond}= \lbrace m \in M \mid \exists g\in G, (gRm,\ g\in X) \rbrace\subseteq M,\nonumber\\ 
Y\subseteq M &\mapsto& Y^{\Diamond}= \lbrace g \in G \mid \exists m\in M, (gRm,\ m\in M) \rbrace\subseteq G,
\label{eq3}
\end{eqnarray}
where, notably, $gRm$ is equivalently $mRg$ since $m\in g^R$ iff $ g \in m^R$. 
\end{df}
Subsequently, it is known that the maps in Eqs (\ref{eq1}-\ref{eq3}) manifest
the following relations, for $X,X_1,X_2 \subseteq G$ and $X^c:=G\backslash X$,
\begin{eqnarray}
X^{III}=X^I,&& 
\begin{array}{c}
X^{\Box\Diamond\Box}=X^{\Box}\\
X^{\Diamond\Box\Diamond}=X^{\Diamond}
\end{array},
\label{eq:triple_o} \\ 
X_1\subseteq X_2\iff (X_2)^I\subseteq (X_1)^I, && 
\begin{array}{c}
X_1\subseteq X_2 \iff (X_1)^{\Box}\subseteq (X_2)^{\Box}\\
X_1\subseteq X_2 
\iff (X_1)^{\Diamond}\subseteq (X_2)^{\Diamond}
\end{array},
\label{eq:derv_order}
\end{eqnarray}
\begin{equation}
 X^{c\Box c}=X^{\Diamond},\quad 
X^{c\Diamond c}=X^{\Box}, \label{eq:complement}
\end{equation}
where the same relations are also applied to $Y,Y_1,Y_2 \subseteq M$ and $Y^c:=M\backslash Y$.
For convenience, let one employ
the notions {\it extent} and {\it intent} \cite{Wi82,GW99} for both FCL and RSL 
such that the two theories can be handled on an equal footing.
\dfn{Consider $X \subseteq G$ and $ Y \subseteq M$ subject to the formal context $F(G,M)$.
\begin{itemize}
\item The 2-tuple $(X,Y)$ is called an FCL concept
if $X^I=Y$ and $Y^I=X$, where 
$X$ is the FCL {\it extent}, $Y$ is the FCL {\it intent} \cite{Wi82,GW99}.
\item
The 2-tuple $(X,Y)$  is called an RSL concept if $X^{\Box}=Y$ and 
$Y^{\Diamond}=X$, where 
$X$ is the RSL {\it extent}, $Y$ is the RSL {\it intent}.
\end{itemize}}
Notably, the 2-tuple $(X,Y)$ with 
$\left \lbrace\begin{smallmatrix}
X^{\Box}=Y\\
Y^{\Diamond}=X
\end{smallmatrix}\right.$ has been referred to as 
{\it object-oriented} RSL concept \cite{DG03,YY04}, in contrast to
the case
of property-oriented concept defined through
 $\left \lbrace\begin{smallmatrix}
X^{\Diamond}=Y\\
Y^{\Box}=X
\end{smallmatrix}\right.$.
However, one will ignore the consideration of $\left \lbrace\begin{smallmatrix}
X^{\Diamond}=Y\\
Y^{\Box}=X
\end{smallmatrix}\right.$ since it
can be derived from 
$\left \lbrace\begin{smallmatrix}
X^{\Box}=Y\\
Y^{\Diamond}=X
\end{smallmatrix}\right.$ 
by means of interchanging the object and attribute.
In addition, as a consequence of Eq.~(\ref{eq:complement}),
\begin{equation}
\left \lbrace\begin{smallmatrix}
X^{\Box}=Y\\
Y^{\Diamond}=X
\end{smallmatrix}\right.\iff \left \lbrace\begin{smallmatrix}
X^{cc\Box c}=Y^c\\
Y^{cc\Diamond c}=X^c
\end{smallmatrix}\right.\iff \left \lbrace\begin{smallmatrix}
(X^c)^{\Diamond}=(Y^c)\\
(Y^c)^{\Box}=(X^c)
\end{smallmatrix}\right.
\label{eq:rsl_duality}, 
\end{equation}
telling \cite{YY04} that 
if $(X_i,Y_i)$ is a concept 
appropriate for $\left \lbrace\begin{smallmatrix}
X_i^{\Box}=Y_i\\
Y_i^{\Diamond}=X_i
\end{smallmatrix}\right.$
then $(X_i^c,Y_i^c)$ is a concept 
appropriate for 
$\left \lbrace\begin{smallmatrix}
X_i^{\Box}=Y_i\\
Y_i^{\Diamond}=X_i
\end{smallmatrix}\right.$
and vice versa.
It is remarkable that
{\bf Definition~2.3} provides fundamental 
ingredients for the original FCL and RSL \cite{GW99,YY04}.
Based on Eq.~(\ref{eq:triple_o}), 
a $2$-tuple that fulfils 
$\left \lbrace\begin{smallmatrix}
X^I=Y\\
Y^I=X
\end{smallmatrix}\right.$
is guaranteed to exist and 
can be written as 
$(X^{II},X^I)$ or $(Y^I,Y^{II})$, while
a $2$-tuple that fulfils 
$\left \lbrace\begin{smallmatrix}
X^{\Box}=Y\\ 
Y^{\Diamond}=X
\end{smallmatrix}\right.$
is guaranteed to be existent and 
can be written as 
$(X^{\Box\Diamond},X^{\Box})$ or $(Y^{\Diamond},Y^{\Diamond\Box})$.
Moreover, the Galois connection composed by the 
two posets $(2^G,\subseteq)$ and $(2^M,\subseteq)$ can 
be accomplished in terms of the {\it concept}s 
by virtue of Eq.~(\ref{eq:derv_order}).

To proceed with the general theory of concept lattice,
it is of crucial importance to clarify the notions contained in formal objects and formal attributes.
Since a formal context 
discerns the objects based on 
the attributes they possess, the objects which possess
the same attributes are grouped into an equivalent class.
\dfn{The discernible object classes in the perspective of
$F(G,M)$ are definite. 
One will call these $F$-distinct subsets of $G$ 
and denote each such class with $D_k$ in the sense 
that $D_k$'s are regarded as 
distinct entities by $F(G,M)$.
The total number of these $F$-distinct
object classes {$n_F$} is less than or equal 
to $|G|$, which is the cardinality of $G$.
In effect, $\forall g\in D_k\ g^R=D_k^I$,
and $D_k^I\neq D_{k^\prime}^I$ iff $D_k\neq D_{k^\prime}$. 
Hence, one has  the collection of $F$-distinct
object classes $G_{/R}=\lbrace D_k\mid k=1\ldots n_F\rbrace$ based on the equivalence relation
\q{$g \approx_R g^\prime$ iff $g^R=g^{\prime R}$}, which means that $\lbrace g, g^\prime\rbrace \subseteq D_k$ for some $D_k \in G_{/R}$.
Likewise, the $F$-{\it co}distinct subsets
are gathered as $G^{/R}=\lbrace G\backslash D_k\mid D_k \in G_{/R}\rbrace$.} 
In contrast to the formal objects, 
the formal attributes 
can also emerge in a {\it composite} manner. 
One is free to construct various {\it new attributes} out of the given attributes by means of 
the standard Boolean operations, thereby forming 
an extended framework for the general consideration.
\dfn{Given a set $M$ of attributes,
the set of {\it generalised attribute}s over $M$ is defined as 
$M^{\ast} =\lbrace b(M_0)\mid M_0\subseteq M \rbrace$,  
where $b$ iterates over all the possibilities of {\it Boolean} functions in which
the concerned operations are 
\q{$\cdot$}~(conjunction), \q{$+$}~(disjunction) and \q{$\neg$}~(negation).
To be concrete, $M^{\ast}$ is meant to denote 
whatever one may construct out of $M$ 
by means of {\it arbitrary compositions} provided by the Boolean functions. In contrast to the members of $M^\ast$, which may be {\it composite}, the attributes in $M$ are then referred to as {\it simple} attributes.}
Intuitively, $M^\ast$ also comprises the set $M$ itself as well as 
the conventional {\it truth} ${\bf 1}$ and {\it falsity} ${\bf 0}$.
Moreover, 
by means of {\it negation} 
one may define several interesting subclasses of $M^\ast$.
\dfn{Let $M$ be certain collection of attributes of interest. 
A {\it conjugate class}
$\Psi_M^i$ is referred to as one of the $2^{|M|}$ possibilities of
$\lbrace \alpha \mid \alpha=\mbox{either}\ a\ \mbox{or}\ \neg a, a \in M\rbrace$.
Accordingly, the set of all {\it conjugate class}es is denoted 
with ${\bf \Psi}_M=\lbrace\Psi_M^1\ldots \Psi_M^{2^{|M|}}\rbrace$. 
The {\it symmetrised attribute set} ${\breve M}$ is employed to extend $M$ 
by means of negation: ${\breve M}=M\cup\lbrace \neg a\mid a\in M \rbrace$.} 
One could also employ the set of generalised attributes as 
$M^{\ast}=({\Psi_M^i})^\ast=\lbrace b(M_0)\mid M_0\subseteq {\Psi_M^i}\rbrace$ with any ${\Psi_M^i}\in {\bf \Psi}_M$.
It is noteworthy that prescribing $M^\ast$,
either by $({\Psi_M^i})^\ast$ or by {\bf Definition 2.5}, 
may encounter 
abundant choices of Boolean function $b$.
Technically, such abundance can be reduced by means of {\it normal forms}, where
one is in particular interested in the {\it conjunctive normal form} ({\bf CNF}) 
and the {\it disjunctive normal form} ({\bf DNF}).
The other choice to suppress the 
redundance
is to consider $M^{\ast}=({\breve M})^{\ast}=\lbrace b_0(M_0)\mid M_0\subseteq {\breve M}\rbrace$, where $b_0$ differs from $b$ by that {\it no further} {negation} is needed (cf. {\bf Definition 2.5}). 
In what follows, the particular significance of the 
{\it generalised attributes over} $M$ 
to the structure of $F(G,M)$
will be presented step by step.

Attributes are considered to be further restrained
through the attribute-object correspondence provided
in the formal context $F(G,M)$.
\dfn{According to {\bf Definition 2.5} any generalised attribute $\mu\in M^\ast$ 
can be regarded as
$\mu(M)$, namely that $\mu$ is a (Boolean) function of $M_0$, say
$\mu=b(M_0)$, for certain $b$ and for some $M_0\subseteq M$.
Subject to $F(G,M)$,
the {\it contextual} (Boolean) function $\mu_F(M)$ 
for any generalised attribute $\mu\in M^\ast$ 
is obtained from the expression of $\mu(M)$ 
by (1)
replacing the concerned attribute $m$ with $m^R$ ($m^R\subseteq G$), (2)
replacing the concerned negative attribute $\neg m$ with $(m^R)^c$, (3)
replacing the conjunction and disjunction operations with $\cap$ and $\cup$, respectively. 
Notably, the 
contextual function $\mu_F(M)$ is essentially a subset of $G$.} 
For the extreme case,  one has \q{$m=m(M)\ \forall m\in M$} such that
$m_F(M)=m^R\subseteq G\ \forall m\in M$.
The contextual function can represent all the information 
contained in the formal context $F(G,M)$ since
the contextual function assigns every attribute, say $m$,
in $M$ with a definite object subset which possesses $m$ in common.
Furthermore, such idea can be extended.
\lem{From $F(G,M)$ one can deduce the {\it extended} formal context $F^\ast(G,M^\ast)$ 
without any additional assumption in the sense that
$\forall \mu\in M^{\ast}$ $\mu^R$ can be determined by means 
of {$\mu^R=\mu_F(M)$}, which is called 
\q{$\mu^R$ {is well defined in} $M^\ast$}.}
\prf{If $\mu^R$ is well defined in $M^\ast$ then
one can achieve the extended formal context $F^\ast(G,M^\ast)$
such that each attribute in $M^\ast$ is equipped with
an object set which possess the attribute in common, similar to the case of $F(G,M)$.
Here, the proof of \q{$\mu^R$ is well defined in $M^\ast$} can be carried out in 
two steps. It is to show that
(1) ${\neg m}^R=({\neg m})_F\equiv (m^R)^c$ ({\bf Definition~2.7}) for $m\in M$ such that $\mu^R$ is well defined in 
${\breve M}$ and (2) both $(\prod_{\mu}\mu)^R$ and $(\sum_{\mu}\mu)^R$
are well defined $\forall \mu\in {\breve M}$.
Once these are done, one will end up with 
\q{$\mu^R$ is well defined in $M^\ast$}
since $M^\ast$ can be obtained as ${\breve M}^\ast$
by means of \q{$+$} and \q{$\cdot$} (see the discussion after {\bf Definition~2.6}).
\begin{itemize}
\item [(1)] 
For $m\in M$ \q{$g\in (\neg m)^R$} can be 
interpreted as \q{$g\not\in m^R$}.
In other words, $g\in G\backslash m^R=(m^R)^c$, 
hence, $(\neg m)^R =(m^R)^c=({\neg m})_F$, which gives rise to {$m^R =((\neg m)^R)^c$} after taking complementarity. 
Moreover, if  
the negated attributes
are also taken into account, the same interpretation implies that
{$(\neg \neg m)^R =((\neg m)^R)^c$}.
This is consistent since {$(\neg \neg m)^R$} coincides with {$m^R$}:
 $gR(\neg \neg m)\iff g\not\in (\neg m)^R\iff g\in ((\neg m)^R)^c\iff g\in ((m^R)^c)^c\iff gRm$. 
\item [(2)] 
Since $\mu^R$ is well defined in ${\breve M}$, 
one may consider that $(\prod_{\mu}\mu)^R
=\lbrace g\mid gR(\prod_{\mu}\mu), g\in G\rbrace= 
\bigcap_{\mu}\mu^R=(\prod_{\mu}\mu)_F$ in ${\breve M}$. 
Likewise, $(\sum_{\mu}\mu)^R
=\lbrace g\mid gR(\sum_{\mu}\mu), g\in G\rbrace= 
\bigcup_{\mu}\mu^R=(\sum_{\mu}\mu)_F$ $\forall \mu \in {\breve M}$.
\end{itemize}
}
%
\dfn{
A {\it contextual} Venn diagram (or Euler diagram) ${\cal V}_M^F$ subject to $F(G,M)$ 
can illustrate the set relations
among the contextual functions of attributes in $M$.
One can 
achieve ${\cal V}_M^F$ 
by encircling the object-set $m^R$ within the object collection $G$, where the object-labellings are ignored. 
In contrast, $M$ may also 
have its own intrinsic logical structure 
in terms of the (conventional) Venn digram ${\cal V}_M$.} 
In practice, there are two types of ordering systems for attributes.
As the first type, \q{$\geq, \leq$} is concerned with 
the intrinsic ordering of 
the entities in $M^\ast$.  
For instance, {$\mu_i\leq \mu_j$}
denotes that the region representing $\mu_i$ are included in   
the region representing $\mu_j$ on ${\cal V}_M$. 
The second type of ordering system \q{$\supseteq,\subseteq$} 
is employed for ${\cal V}_M^F$
because $\mu^R$s 
are in fact object sets.
Notably, preserving the object labellings in 
the contextual Venn digram 
will restore all the information content of the formal context.
The contextual Venn diagram with explicit object labellings
turns out to be an alternative representation for the formal context.
Once ${\cal V}_M^F$ is given, {\bf Lemma~2.8} becomes a rather
intuitive result since one may 
read off every $\mu^R$ for $\mu\in M^\ast$.
The other point is that 
the formal context
does not per se suggest  
any intrinsic relation among members in $M$.
Attributes in $M$ can be basically independent,
hence, any two attributes have an intersection 
on the corresponding Venn diagram ({${\cal V}_M={\cal V}_M^0$}).
However, it is also allowable that 
one imposes additional constraints on $M$
by removing some of the disjoint regions from ${\cal V}_M^0$.
\lem{Subject to the formal context $F(G,M),\ \forall \mu\forall\mu_i\forall\mu_j\in M^\ast$,
\begin{itemize}
\item
$\mu^R\neq \emptyset\implies \mu\neq {\bf 0}$,
\item
$\mu_i=\mu_j \implies\ \mu_i^R=\mu_j^R$ but $\mu_i^R\neq\mu_j^R\implies\ \mu_i\neq\mu_j$,
\item
$\mu_i<\mu_j  \implies\ \mu_i^R\subseteq\mu_j^R$ but $\mu_i^R\subset \mu_j^R \implies\ \mu_i\not> \mu_j$.
\end{itemize}}
\prf{
\begin{itemize}
\item
If $\mu={\bf 0}$ but $\mu^R=X\neq \emptyset$ then the existence of the object set $X$ is contradictory. 
\item
$\mu_i=\mu_j \implies\ \mu_i^R=\mu_j^R$ since from 
{$\mu_i(M)=\mu_j(M)$} it follows that 
$(\mu_i)_F(M)=(\mu_j)_F(M)$ ({\bf Lemma~2.8}). 
If $\mu_i^R\neq\mu_j^R$ then 
$\mu_i^{Rc}\cap\mu_j^R\neq \emptyset$ and/or $\mu_i^R\cap\mu_j^{Rc}\neq \emptyset$ 
which means that {$(\neg\mu_i\cdot\mu_j)^R\neq \emptyset$ and/or $(\mu_i\cdot\neg\mu_j)^R\neq \emptyset$}. Consequently, 
$(\neg\mu_i\cdot\mu_j)\neq {\bf 0}$ and/or $(\mu_i\cdot\neg\mu_j)\neq {\bf 0}$, which contradicts $\mu_i=\mu_j$. 
\item
For $\mu_i<\mu_j$ one may assume $\mu_j=\mu_i+\beta$ in which $\beta\in M^\ast$.
Then, $\mu_j^R=\mu_i^R\cup\beta^R\supseteq \mu_i^R$ according to {\bf Lemma~2.8}.
Moreover, if $\mu_i^R\subset \mu_j^R$ but $\mu_i> \mu_j$, contraction occurs 
since {$\mu_i> \mu_j$} implies {$\mu_i^R\supseteq\mu_j^R$} which denies {$\mu_i^R\subset \mu_j^R$}.
Thus, $\mu_i^R\subset \mu_j^R \implies\ \mu_i\not> \mu_j$. 
\end{itemize}}
The above shows that the {\it distinction} 
between $\mu_i$ and $\mu_j$ is stronger than the pair $(\mu_i^R, \mu_j^R)$.
More concretely, the number of disjoint regions on ${\cal V}_M^0$ is $2^{|M|}$,
since ${\cal V}_M^F$ is less discernible than ${\cal V}_M^0$,
thus $2^{|M|}\geq n_F$ 
\footnote{
The formal context with $2^{|M|}=n_F$ will be named {\it degenerate} formal context \cite{LLJD12-2}, 
because the {\it Grsp} is then equivalent to {\it Gfcp} ({\bf Proposition~3.4}) 
at each general extent. 
Although the {degenerate formal context} is 
not a practical example for the problem of categorisation,
it provides a very useful theoretical tool
for inspecting the {\bf GCL} structure in every detail \cite{LLJD12-3}\cite{LLJD12-4}.}.
\dfn{The {\it contextual equivalent class of attribute} subject to $F(G,M)$
is an attribute-set defined in the following manner.
Given $\nu\in M^\ast$, one may collect all the attributes which have the same contextual function ({\bf Definition~2.7}) with $\nu$ as  
$\lbrace \nu\rbrace_F=\lbrace \mu\mid \mu^R=\nu^R,\ \mu \in M^\ast \rbrace$.
Accordingly, since $\nu^R\subseteq G$, 
the idea can be employed with respect to definite object-set in the sense that 
$[X]_F=\lbrace \mu\mid \mu^R=X,\ \mu \in M^\ast \rbrace$ for $X\subseteq G$.}
It is noteworthy that the construction $(X,[X]_F)$
is in fact the precursor of the {\it concept} on the {\bf GCL}.
Here, the remaining condition lies on determining
whether an arbitrary subset of $G$ can form a node of the {\bf GCL}. 
Remarkably, 
a subset $X$ of $G$ with {\it trivial} contextual equivalent class, i.e. 
$[X]_F=\emptyset$, 
can hardly participate in the categorization provided by the formal context 
since no attribute can be found to label it.
The point is that any of the $F$-distinct subsets of $G$ ({\bf Definition~2.7}),
say $D_k$, 
is readily a smallest possible subset equipped with non-trivial contextual equivalent attributes.
In other words, the $F$-distinct subsets correspond to the smallest regions 
on the contextual Venn diagram ({\bf Definition~2.9}), hence, 
there exists no attribute, say $\mu$, 
whose contextual function $\mu^R$ exactly encloses a region which is smaller than the region marked by $D_k$.
Consequently, for $X$ that can be written as the union of $F$-distinct subsets
($X=\bigcup_{k_X} D_{k_X}$) one 
has non-trivial $[X]_F$, otherwise $[X]_F=\emptyset$.
 
A further
instructive notion is the {\it smallest yet non-trivial} 
constituents of the generalised attribute set, for which one may develop two 
independent {\it non-trivial minimisation} aspects.
Firstly,  
one may speak of
the minimal attributes of $M^\ast$ which are {\it not} the 
Falsity {${\bf 0}$} and thus correspond to the smallest regions on the Venn diagram ${\cal V}_M$.
Secondly, subject to $F(G,M)$, there are classes of {\it irreducible}
attributes 
that determines the
object classes which participate in the categorisation
in a {\it least possible} manner.
Here, the first non-trivial minimisation aspect 
is a prevalent issue which deserves clarification. 
\lem{
Given an attribute-set $M$, one may define
the set of {\it non-trivial infima} 
{$b_{inf}(M^\ast):=\lbrace \tau\in M^\ast\mid \tau\succ_M {\bf 0} \rbrace$}
and the set of {\it non-trivial suprema} 
{$b_{sup}(M^\ast):=\lbrace \tau\in M^\ast\mid {\bf 1}\succ_M \tau\rbrace$}
for the corresponding generalised attribute-set $M^\ast$.
In the case of
{$\mu\succ_M\nu$}, one denotes that $\mu>\nu$ without any attribute $\chi\in M^\ast$ which fulfils {$\mu>\chi>\nu$}.
Both $b_{inf}(M^\ast)$ and $b_{sup}(M^\ast)$
can be expressed in terms of ${\bf \Psi}_M$ ({\bf Definition~2.6}):
\begin{itemize}
\item
$b_{inf}(M^\ast)=\left\lbrace \prod\Psi_M^j\mid \Psi_M^j\in  {\bf \Psi}_M\right\rbrace$,
$\prod\Psi_M^j$ is called $M^\ast$-atom, which is considered as an {\it atom} for $M^\ast$. 
\item
$b_{sup}(M^\ast)=\left\lbrace \sum\Psi_M^j\mid \Psi_M^j\in  {\bf \Psi}_M\right\rbrace$,
$\sum\Psi_A^j$ is called $M^\ast$-coatom.
\end{itemize}
}
\prf{
One may rewrite
$\tau\succ_M {\bf 0}$
as \q{$\tau\cdot \beta\in \left\lbrace \tau,{\bf 0}\right\rbrace\ \forall  \beta \in M^\ast$}.
\begin{itemize}
\item
For any $\beta \in M^\ast$ 
one may write down the {\bf DNF} 
as $\beta=\sum_{i_\beta} \prod\Psi_M^{i_\beta}$, where the range of $i_\beta$ is up to the given $\beta$.
Hence, $\forall j\forall \beta\ \beta\cdot \prod\Psi_M^j
=\left(\sum_{i_\beta} \prod\Psi_M^{i_\beta}\right)\cdot \prod\Psi_M^j  
\in \left\lbrace \prod\Psi_M^j,{\bf 0}\right\rbrace,\ \therefore\ \forall j\ \prod\Psi_M^j\in b_{inf}(M^\ast)$.
On the other hand, assume $\exists \chi \in b_{inf}(M^\ast)$ but 
$\chi\neq \prod\Psi_M^j$ for $\Psi_M^j\in {\bf \Psi}_M$.
However, attribute
$\chi$ may take the form $\sum_{i_\chi} \prod\Psi_M^{i_\chi}$ in {\bf DNF},
which apparently implies $\chi\not \succ_M {\bf 0}$.
One concludes then $b_{inf}(M^\ast)=\left\lbrace \prod\Psi_M^j\mid \Psi_M^j\in  {\bf \Psi}_M\right\rbrace$.
\item
$\beta+\sum\Psi_M^j\equiv \beta\cdot \neg(\sum\Psi_M^j)+\sum\Psi_M^j$
$\forall j\forall \beta$.
Subsequently, $\beta\cdot \neg(\sum\Psi_M^j)\in \left\lbrace \neg(\sum\Psi_M^j),{\bf 0}\right\rbrace$ based on the above result since 
$\neg(\sum\Psi_M^j)$ can be identified as $\prod\Psi_M^{j^\prime}$ for some $\Psi_M^{j^\prime}\in {\bf \Psi}_M$, which implies
that $\beta+\sum\Psi_M^j
\in 
\left\lbrace {\bf 1},\sum\Psi_M^j\right\rbrace$, i.e. 
$\sum\Psi_M^j\in b_{sup}(M^\ast)$.
On the other hand, assume $\exists \chi \in b_{sup}(M^\ast)$ but 
$\chi\neq \sum\Psi_M^j$ for $\Psi_M^j\in {\bf \Psi}_M$.
However, attribute
$\chi$ may take the form $\prod_{i_\chi} \sum\Psi_M^{i_\chi}$ in {\bf CNF},
which apparently implies $\chi\not\prec_M 1$.
Therefore, $b_{sup}(M^\ast)=\left\lbrace \sum\Psi_M^j\mid \Psi_M^j\in  {\bf \Psi}_M\right\rbrace$. 
\end{itemize}}
\crl{
Whenever there are no intrinsic conditions pre-imposed on the attributes
in $M$, the cardinalities are as follows.
\begin{itemize}
\item
The numbers of non-trivial suprema and infima for $M^\ast $ are equivalent: 
$|b_{sup}(M^\ast)|=|b_{inf}(M^\ast)|=2^{|M|}$. 
\item
There are $2^{2^{|M|}}$ distinct {\it generalised} attributes in $M^\ast$, i.e., $|M^\ast|=2^{2^{|M|}}$. 
\end{itemize}
}
\prf{Whenever there are no intrinsic conditions pre-imposed on the attributes
in $M$, the elements of $b_{inf}(M^\ast)$ can be represented by the $2^{|M|}$ disjoint
regions on ${\cal V}_M^0$.
\begin{itemize}
\item 
Alternatively, by {\bf Definition~2.6} one may compute that
$|b_{inf}(M^\ast)|=|{\bf \Psi}_M|=2^{|M|}$. 
The elements of $b_{sup}(M^\ast)$
amount to the complementary parts of the disjoint regions on ${\cal V}_M^0$,
which demonstrates an one-to-one correspondence; 
$|b_{sup}(M^\ast)|=|b_{inf}(M^\ast)|$. 
\item
Denote the members in $b_{inf}(M^\ast)$ with $b_k$ where $1\leq k\leq 2^{|M|}$.
An attribute in $M^\ast$, say $\mu$, 
can take the expression
$\sum_{k_\mu} b_{k_\mu}$ in {\bf DNF}. 
In other words, the attributes 
in $M^\ast$ may 
be implemented by $\sum_{k\in M_1} b_k$ where $M_1$
ranges over all the possibilities for 
$M_1\subseteq b_{inf}(M^\ast)$.  
Counting all these possibilities is equivalent to enumerating the number of 
the power sets 
of a set with $2^{|M|}$ members, i.e., $|M^\ast|=2^{2^{|M|}}$.  
\end{itemize}
}
The second non-trivial minimisation aspect mentioned above is related to the formal context.    
\dfn{A generalised attribute $\mu\in_{F+} M^\ast$ composed of 
the {\it disjunction} of attributes in ${\breve M}$ ({\bf Definition~2.6})
is called an {\it irreducible disjunction}  
subject to $F(G,M)$ if 
{eliminating any term in the disjunction from $\mu$}, which results in $\mu^\prime$, always
causes {$(\mu^\prime)^R\subset \mu^R$}. 
Accordingly, an $X$-{\it irreducible disjunction class} can be defined as
{$[X^{+}]_F=\lbrace \mu\in [X]_F\mid \mu\in_{F+} M^\ast \rbrace$}, 
see {\bf Definition~2.11}.
On the other hand,
a generalised attribute $\mu\in_{F\times} M^\ast$ composed of 
the {\it conjunction} of attributes in ${\breve M}$ 
is called an {\it irreducible conjunction} in $M^\ast$ 
subject to $F(G,M)$ if 
{eliminating any term in the conjunction from $\mu$}, which results in $\mu^\prime$, always
causes {$(\mu^\prime)^R\supset \mu^R$}. 
The $X$-{\it irreducible conjunction class} is given as 
{$[X^{\times}]_F:=\lbrace \mu\in [X]_F\mid \mu\in_{F\times} M^\ast\rbrace$}.} 
In practice, {$\mu\in_{F+} M^\ast$} can be written in terms of 
{$\mu=\sum_{\alpha\in \psi}\alpha$} 
for some $\psi\subseteq \Psi_M^j$ ({\bf Definition~2.6}) in which $\mu^R\neq \left(\sum_{\alpha\in \psi\backslash \lbrace\alpha_0\rbrace}\alpha\right)^R\ \forall  \alpha_0\in \psi$. 
Likewise, 
{$\mu\in_{F\times} M^\ast$} can be written in terms of
{$\mu=\prod_{\alpha\in \psi}\alpha$} for some $\psi\subseteq  \Psi_M^j$ 
where $\mu^R\neq \left(\prod_{\alpha\in \psi\backslash \lbrace\alpha_0\rbrace }\alpha\right)^R\ \forall  \alpha_0\in \psi$.
Generally speaking,
$[X^{+}]_F\cup [X^{\times}]_F\subseteq [X]_F$ and
$[X^{+}]_F\cap [X^{\times}]_F\neq\emptyset$.
The other useful expressions are
\begin{eqnarray}
\frac{[X_0^{\times}]_F}{[X_i^{\times}]_F}&:=&
\lrbr{\mu\in [X_0^{\times}]_F\mid \mu\neq \nu_1\cdot\nu_2,\ 
\nu_1 \in  [X_i^{\times}]_F\ \mbox{or}\ 
\nu_2 \in [X_i^{\times}]_F
}\subseteq [X_0^{\times}]_F,\nonumber\\
\frac{[X_0^{+}]_F}{[X_j^{+}]_F}&:=&\lrbr{\mu^\prime\in [X_0^{+}]_F\mid \mu^\prime\neq \nu_1^\prime+\nu_2^\prime,\ 
\nu_1^\prime \in  [X_j^{+}]_F\ \mbox{or}\ 
\nu_2^\prime \in  [X_j^{+}]_F
}\subseteq [X_0^{+}]_F.\label{eq:irred_subtract}
\end{eqnarray}
Note that $\frac{[X_0^{\times}]_F}{[X_i^{\times}]_F}\neq [X_0^{\times}]_F$ only when $X_i\supset X_0$ and 
$\frac{[X_0^{+}]_F}{[X_j^{+}]_F}\neq [X_0^{+}]_F$
only when $X_j\subset X_0$.  
What follows is mainly concerned with regaining the FCL and RSL nodes on the {\bf GCL}.
\lem{ 
\begin{itemize}
\item 
$\forall {\breve m}\in {\breve M}\quad {\breve m}\in [({\breve m}^R)^{+}]_F \cap [({\breve m}^R)^{\times}]_F$ and, furthermore, $\forall X_i\forall X_j\neq {\breve m}^R\ ~{\breve m}\in 
\frac{[({\breve m}^R)^{+}]_F}{[X_i^{+}]_F}
\cap \frac{[({\breve m}^R)^{\times}]_F}{[X_j^{\times}]_F}$.
\item $\forall  \mu\in M^\ast\quad 
\mu\in [X^{+}]_F\implies\neg\mu\in [\left(X^c\right)^\times]_F$ and
$\mu\in [X^{\times}]_F\implies\neg\mu\in [\left(X^c\right)^+]_F$.
\end{itemize}
}
\prf{
\begin{itemize}
\item
Any attribute ${\breve m}\in {\breve M}$ ({\bf Definition~2.6})
is non-composite and 
can be thus regarded as 
both an one-term disjunction and an one-term conjunction. 
Therefore, ${\breve m}$ is simultaneously an ${\breve m}^R$-{\it irreducible disjunction} and an ${\breve m}^R$-{\it irreducible conjunction}. 
Moreover, ${\breve m}\neq \nu_1\cdot\nu_2$ in which $\nu_1 \in  [X_i^{\times}]_F\ \mbox{or}\ 
\nu_2 \in [X_i^{\times}]_F$ and 
${\breve m}\neq \nu_1^\prime+\nu_2^\prime$ in which $\nu_1^\prime \in  [X_j^{+}]_F\ \mbox{or}\ \nu_2^\prime \in  [X_j^{+}]_F$.
\item
For $\mu\in [X^{+}]_F$, let $\mu=\sum_{\alpha\in \psi}\alpha$ where $\psi\subseteq \Psi_M^j$ for some
$\Psi_M^j\in {\bf \Psi}_M$. Accordingly, $\mu^R=X$ but 
$\forall \alpha_1 \in \psi$ $\left(\sum_{\alpha\in \psi\backslash \lbrace \alpha_1\rbrace}\alpha\right)^R\ \neq X$, which implies
that $(\neg \mu)^R=X^c$ but 
$\forall \alpha_1 \in \psi$ $\left(\sum_{\alpha\in \psi\backslash \lbrace \alpha_1\rbrace}\alpha\right)^{Rc}
\neq X^c
$.
It then turns out that 
\[
(\neg \mu)^R=\lrp{\prod_{\alpha\in \psi}\neg\alpha}^R
=\lrp{\prod_{\alpha\in \psi^\prime}\alpha}^R=X^c\quad \mbox{but}\quad \forall \neg \alpha_1 \in \psi^\prime\ 
\left(\prod_{\alpha\in \psi^\prime\backslash \lbrace \neg \alpha_1\rbrace}\alpha\right)^R\neq X^c,
\]
where
$\psi^\prime=\lbrace \neg\alpha\mid \alpha\in \psi\rbrace\subseteq \Psi_M^{j^\prime}$ with $\Psi_M^{j^\prime}=\lbrace \neg\alpha\mid \alpha\in \Psi_M^j\rbrace\in {\bf \Psi}_M$.
Therefore, $\mu\in [X^{+}]_F \implies \neg\mu\in [\left(X^c\right)^\times]_F$.
Similarly, starting with $\mu^\prime=\sum_{\alpha\in \psi}\alpha\in [X^{\times}]_F$, one will end up with $\mu^\prime\in [X^{\times}]_F\implies\neg\mu^\prime\in [\left(X^c\right)^+]_F$.
\end{itemize}
}

\section{obtaining the general concept lattice}\label{three}
%
The general theory of concept lattice is capable of
providing the categorisation for the formal objects $G$ according to members of $M^\ast$.
First of all, one observes that
one could have replaced 
both FCL and RSL intents in the original constructions 
in terms of {\it single composite} attributes
without altering their information contents.
\lem{
Let $X\subseteq G$, $Y\subseteq M$ for the formal context $F(G,M)$. 
The concepts 
for FCL and RSL ({\bf Definition 2.3}) can be both expressed in terms of $(X,\mu)$ in the following manner.
\begin{itemize}
\item
For FCL, use $\mu=\prod Y$, where $\mu$ is referred to as a 
{formal concept property} ({\it fcp}).
Imposing $\lsmatr{X^I=Y\\
Y^I=X}$
on $(X,Y)$
is indistinguishable from imposing 
$\lsmatr{\prod X^I=\mu\\
\mu^R=X}$ on $(X,\mu)$.  
\item
For RSL, use $\mu=\sum Y$, where $\mu$ is referred to as a rough set property ({\it rsp}).
Imposing $\lsmatr{X^{\Box}=Y\\
Y^{\Diamond}=X}$
on $(X,Y)$
is indistinguishable from imposing 
$\lsmatr{\sum X^\Box=\mu\\
\mu^R=X}$
on $(X,\mu)$. 
\end{itemize}
}
\prf{
\begin{itemize}
\item
$X^I=Y \implies \prod X^I=\prod Y$. 
Reversely, since both $X^I$ and $Y$ consist of distinct members in $M$,
$\prod X^I=\prod Y \implies X^I=Y$. Thus, $X^I=Y \iff \prod X^I=\prod Y$and it is appropriate to employ $\mu=\prod Y$.  
Moreover, by Eq. (\ref{eq1}) $\mu^R=(\prod Y)^R=\bigcap_{y\in Y}y^R=\lbrace g\in G\mid\ gRy,\ \forall y \in Y\rbrace=Y^I$. Consequently, $\mu^R=X\iff Y^I=X$. 
\item 
$X^{\Box}=Y \implies \sum X^\Box=\sum Y$. 
Reversely, since both $X^{\Box}$ and $Y$ consist of distinct members in $M$,
$\sum X^\Box=\sum Y \implies X^{\Box}=Y$. Thus, $X^{\Box}=Y \iff \sum X^\Box=\sum Y$ and it is appropriate to employ $\mu=\sum Y$.  
Moreover, by Eq. (\ref{eq2}) and (\ref{eq3}) 
$\mu^R=(\sum Y)^R=\bigcup_{y\in Y}y^R=\lbrace g\in G\mid\ \exists y\in M,\ (gRy,\ y\in Y)\rbrace=Y^\Diamond$. Consequently, $\mu^R=X\iff Y^\Diamond=X$. 
\end{itemize}
}
Since the above proof has invoked
some of variants of derivation operators (see Eq.~(\ref{eq1})-(\ref{eq3})),  
it is worthwhile to mention about all of such variants:
\begin{eqnarray}
 X^I=\lbrace m\in M \mid gRm,\ \forall g \in X \rbrace&\equiv& \lrbr{m\in M\mid m^R\supseteq X}=\bigcap_{x\in X}x^R,\nonumber\\
 X^{\Box}=\lbrace m\in M \mid \forall g \in G, gRm \implies g\in X \rbrace
 &\equiv& \lrbr{m\in M\mid m^R\subseteq X},\nonumber\\
 X^{\Diamond}=\lbrace m \in M \mid \exists g\in G, (gRm,\ g\in X) \rbrace
&\equiv&\bigcup_{x\in X}x^R,  \label{eq:atderiv}
\end{eqnarray}
where it is intuitive that interchanging the role of $X$ and $Y$
also gives rise to the other set of derivation operators.
Another point is that there are particular object-classes which are fundamental in both the FCL and the RSL.
\prp{
Given a formal context $F(G,M)$, 
$\forall m \in M$ the object-set $m^R$ 
is a {\it  common} extent for FCL and RSL.}
\prf{
$(m^R,(m^R)^I)$ coincides with $(Y^I,Y^{II})$ and
$(m^R,(m^R)^\Box)$ coincides with $(Y^{\Diamond},Y^{\Diamond\Box})$, where in both cases
$Y$ is identified as $\lbrace m\rbrace$.}

Note that the result of {\bf Proposition~3.2} is rather {\it intriguing}, albeit true. Since the formal context conventionally develops a table, 
{\bf Proposition~3.2} in fact entails that 
the object-class $m^R$ is 
recognised as an {\it extent} for both the FCL and RSL 
{\it just} because $m$ is listed in the table. 
Now, assume that there is a formal concept {$(X_1,Y_1)$} 
for FCL subject to $F(G,M)$. 
Then, by virtue of {\bf Proposition~3.1} one has {$\mu_1^R=X_1$}
with {$\mu_1=\prod Y_1$}. 
Consider $F^\prime(G,M\cup \lbrace \mu_1\rbrace)$ 
which is obtained by explicitly including the given $(X_1,\mu_1)$ correspondence 
as an additional column in $F(G,M)$. 
By {\bf Lemma~2.8} 
{$\mu_1^R=X_1$} is something deducible from 
$F(G,M)$, thus, $F^\prime(G,M\cup \lbrace \mu_1\rbrace)$ should {\it not} 
provide different concept lattices.
However, $\mu_1$ is now listed in the new table {$F^\prime(G,M\cup \lbrace \mu_1\rbrace)$}, which implies 
that {$X_1$ must also be an RSL-extent when it is known to be 
an FCL extent}.
This is {\it denied} by
the original FCL and RSL (see the comparison in Ref.~\cite{YY04}),
therefore, it is unnatural 
to neglect the role the composite attribute may play 
just because it is {\it not simple} ({\bf Definition~2.5}).
Hereafter, the general theory of concept lattice will proceed 
in a different way from the original ones
by attempting a democratic consideration for all the members in $M^\ast$.

The concept lattice subject to $F(G,M)$
inevitably makes reference on   
its accompanied extended formal context, i.e. $F^\ast(G,M^\ast)$.
Since $\mu^R$ is well defined $\forall \mu \in M^\ast$ ({\bf Lemma 2.8}),
it is straightforward to manipulate 
$F^\ast(G,M^\ast)$ in parallel to the conventional formal context.
\dfn{
Following from {\bf Definition~2.2},
the derivation operators appropriate for $F^\ast(G,M^\ast)$
are given as  
\begin{eqnarray}
X\subseteq G &\mapsto& X^{I^\ast}=\left\lbrace \mu\in M^\ast \mid gR\mu,\ g \in X \right\rbrace\subseteq M^\ast,
\nonumber\\ 
Y\subseteq M^\ast &\mapsto& Y^{I^\ast}=\lbrace g\in G \mid gR\mu,\ \forall m \in Y \rbrace\subseteq G,\nonumber\\ 
X\subseteq G &\mapsto& X^{\Box^\ast}=\left\lbrace \mu\in M^\ast \mid \forall g \in G,\ gR\mu \implies g\in X \right\rbrace\subseteq M^\ast,\nonumber\\
Y\subseteq M^\ast &\mapsto& Y^{\Box^\ast}=\lbrace g\in G \mid \forall m \in M^\ast,\ gR\mu \implies \mu\in M^\ast \rbrace\subseteq G,
\nonumber\\ 
X\subseteq G &\mapsto& X^{\Diamond^\ast}=\lbrace \mu \in M^\ast \mid \exists g\in G,\ (gR\mu,\ g\in X) \rbrace\subseteq M^\ast,\nonumber\\
Y\subseteq M^\ast &\mapsto& Y^{\Diamond^\ast}= \lbrace g \in G \mid \exists \mu\in M^\ast,\ (gR\mu,\ \mu\in M^\ast) \rbrace\subseteq G.\nonumber
\end{eqnarray}
}
Obviously, 
Eq. (\ref{eq:triple_o})
to (\ref{eq:complement}) remain valid after the substitutions 
\q{$I\ \mbox{by}\ I^\ast$, $\Box\ \mbox{by}\ \Box^\ast$, $\Diamond\ \mbox{by}\ \Diamond^\ast$}. 
\prp{
Subject to $F^\ast(G,M^\ast)$,
the {\it general concept} $(X,\rho(X),\eta(X))$
can be obtained 
as a generalisation of the original FCL and RSL concepts ({\bf Lemma 3.1}) as follows.
If $\rho(X)=\sum X^{\Box^\ast},\ \eta(X)=\prod X^{I^\ast}$ and $(\rho(X))^R=(\eta(X))^R=X$ then
$X$ is called the {\it general extent}, where
the allowable general extents are 
collected as {$E_F:=\lbrace X\subset G\mid [X]_F\neq \emptyset\rbrace=\left\lbrace \mu^R\mid\ \mu\in M^\ast\right\rbrace$}.
In addition,
$\rho(X)$ is the {\it generalised rough set property}~({\it Grsp})
and $\eta(X)$ is the {\it generalised formal concept property}~({\it Gfcp}).} 
\prf{
The {general concept} here
is the consequence of applying {\bf Proposition~3.2} to $F^\ast(G,M^\ast)$, 
which is telling 
that $\forall \mu\in M^\ast\ \mu^R$ is a common extent
of FCL and RSL in the perspective of $F^\ast(G,M^\ast)$.
In practice, $(X,\rho(X),\eta(X))$ represents the dual use of $(X,\mu)$ in {\bf Proposition~3.1} generalised in the sense of {\bf Definition~3.3}.
Accordingly, 
$\left\lbrace \begin{smallmatrix}
(X,\rho(X))=(X, \sum X^{\Box^\ast})\\
(X,\eta(X))=(X, \prod X^{I^\ast})
\end{smallmatrix}\right.$ as well as $(\rho(X))^R=X=(\eta(X))^R$.}
The {\it general extent} can {$X$} then take the form of
{$\mu^R$} with $\mu=\rho(X)$ or $\mu=\eta(X)$, thereby being referred to as an
object class with non-trivial contextual equivalent attributes
({\bf Definition~2.11}).
\prp{
Let {$E_F=\lbrace X\subset G\mid [X]_F\neq \emptyset\rbrace\equiv\left\lbrace \mu^R\mid\ \mu\in M^\ast\right\rbrace$} denote the collection of
all the object classes with non-trivial contextual equivalent classes of attributes.
\begin{itemize}
\item
The set $E_F$ is the sigma algebra 
$\sigma (\left\lbrace m^R\mid m\in M\right\rbrace)$.
\item
$E_F=\sigma({G_{/R}})=\lrbr{\bigcup_{D\in E^0} D\mid E^0\subseteq G_{/R}}$
is only concerned with 
{\it union}s of the members in ${G_{/R}}$ and $|E_F|=2^{n_F}$.
\item
Given $\lrbr{X_1,X_2.\ldots}\subseteq E_F$ then 
$\sigma(\lrbr{X_1,X_2.\ldots})\subseteq E_F$.
\end{itemize}
}
\prf{
The significance of the sigma algebra generated by {\it a collection of sets} is 
that it {\it exhaust}s whatever constructable out of the collection by means of 
iterating various steps of set operations such as {\it intersection, union and complementarity}.
The sigma algebra is essentially closed under {intersection, union and complementarity}.
\begin{itemize}
\item
$E_F=\lbrace \mu^R\mid\ \mu\in M^\ast\rbrace$ ({\bf Proposition~3.4}), 
where 
$\mu^R=\mu_F\ \forall \mu\in M^\ast$ is  well defined according to {\bf Lemma~2.8}.
Hence, $E_F$ can be obtained from $\left\lbrace m^R\mid m\in M\right\rbrace$ by means of {intersection, union and complementarity} ({\bf Definition~2.7}).
Moreover, upon representing the member of $E_F$ as $\mu_i^R$ for some $\mu_i\in M^\ast$, the closure relations are as follows. \\
{$\mu_1^R\cap \mu_2^R=(\mu_1\cdot\mu_2)^R\in E_F$ since $\mu_1\cdot\mu_2\in M^\ast$},
\\ 
{$\mu_1^R\cup \mu_2^R=(\mu_1+\mu_2)^R\in E_F$ since $\mu_1+\mu_2\in M^\ast$},\\
{$(\mu_3^R)^c=(\neg \mu_3)^R\in E_F$ since $\neg\mu_3\in M^\ast$}.\\
Therefore, $E_F$ is closed under {intersection, union and complementarity}. 
\item
Firstly, $E_F\supseteq G_{/R}$ 
in that $\forall D_k\in G_{/R}\ \exists \mu_k$ s.t. $\mu_k^R=D_k$.
For instance, 
{$\mu_k^R=D_k$} is satisfied by taking {$\mu_k=\prod \left\lbrace m\mid m\in D_k^I\subseteq M\right\rbrace\cdot \prod \left\lbrace \neg m\mid 
m\in M\backslash D_k^I\right\rbrace$}.
Moreover, $E_F\supseteq \sigma (G_{/R})$ due to the fact that $E_F$ is itself a sigma algebra.
On the other hand, 
$\sigma (G_{/R})\supseteq \left\lbrace m^R\mid m\in M\right\rbrace$ 
since, e.g., $X_0:=m^R$ can be expressed in terms of $\bigcup_{k_{X_0}}D_{k_{X_0}}$. 
Likewise, $\sigma (G_{/R})\supseteq \sigma (\left\lbrace m^R\mid m\in M\right\rbrace)=E_F$ since
$\sigma (G_{/R})$ is a sigma algebra.
Therefore, $E_F=\sigma (G_{/R})$. 
Secondly, $\sigma (G_{/R})=\lrbr{\bigcup_{D\in E^0} D\mid E^0\subseteq G_{/R}}$
because $G_{/R}$ only comprises the disjoint members 
\q{$D_1,D_2\ldots,D_{n_F}$} where $\bigcup_{k=1}^{n_F}D_k=G$.
Accordingly, one obtains that {$|E_F|=|\sigma (G_{/R})|=2^{n_F}$}. 
\item
$\sigma(\lrbr{X_1,X_2.\ldots})$ is the 
smallest collection that is closed under {intersection, union and complementarity}
and contains $\lrbr{X_1,X_2.\ldots}$, therefore,
$\lrbr{X_1,X_2.\ldots}\subseteq \sigma(\lrbr{X_1,X_2.\ldots})\subseteq E_F$
because $E_F$ is a sigma algebra. 
\end{itemize}
}
\prp{
The requirement for the general concept $(X,\rho(X),\eta(X))$ 
in {\bf Proposition~3.4} is satisfied if $X\in E_F$.\\*
$E_F$ is in effect the full collection of general extents for the {\bf GCL}.
\begin{itemize}
\item
$\forall X\in E_F,\ \rho(X) =\sum_{X_0\subseteq X}\left(\sum [X_0]_F\right)$ and $\eta (X)=\prod_{X_0\supseteq X}\left(\prod [X_0]_F\right)$
where in both cases $X_0\in E_F$.
\item
Equivalently, $\rho (X)=\sum [X]_F$ and $\eta (X)=\prod [X]_F$.
\end{itemize}
}
\prf{
It is remarkable for any subset {$O\subseteq G$} that 
{$O\in E_F$} implies $[O]_F\neq \emptyset$ otherwise $[O]_F=\emptyset$. 
In what follows, $X$ is reserved for the use of {$X\in E_F$} such that 
 $\sum [X]_F$ and $\prod [X]_F$ can be defined. 
\begin{itemize}
\item
According to {\bf Definition~3.3}, \\ 
$\rho(X)=\sum X^{\Box^\ast}=\sum\left\lbrace \mu\in M^\ast \mid \forall g \in G,\ gR\mu \implies g\in X \right\rbrace=
\sum\left(\bigcup_{O\subseteq X}[O]_F\right)=\sum_{X_0\subseteq X}\left(\sum [X_0]_F\right)$,\\ 
$\therefore\ \rho(X)^R=\left(\sum_{X_0\subseteq X}\left(\sum [X_0]_F\right)\right)^R=\bigcup_{X_0\subseteq X}(\bigcup_{\mu \in [X_0]_F}\mu^R)=X$, which
satisfies {\bf Proposition~3.4}, \\
$\eta(X)=\prod X^{I^\ast}=\prod\left\lbrace \mu\in M^\ast \mid gR\mu,\ g \in X \right\rbrace=\prod\left(\bigcup_{O\supseteq X}[O]_F\right)=\prod_{X_0\supseteq X}\left(\prod [X_0]_F\right)$,\\
$\therefore\ \eta(X)^R=\left(\prod_{X_0\supseteq X}\left(\prod [X_0]_F\right)\right)^R=\bigcap_{X_0\supseteq X}(\bigcap_{\mu \in [X_0]_F}\mu^R)=X$, ditto. 
\item
If $\mu \in [X]_F$ then {$\mu+\sum_{X_0\subset X}[X_0]_F \in [X]_F$} as well as {$\mu\cdot\prod_{X_0\supset X}[X_0]_F \in [X]_F$}.\\
Therefore, $\sum [X]_F=\sum_{\mu\in [X]_F} \mu=\sum_{\mu\in [X]_F}\left( \mu+\sum_{X_0\subset X}[X_0]_F\right)=\sum_{X_0\subseteq X}\left(\sum [X_0]_F\right)$\\ 
and $\prod [X]_F=\prod_{\mu\in [X]_F}\mu=\prod_{\mu\in [X]_F}\left(\mu\cdot\prod_{X_0\supset X}[X_0]_F\right)
=\prod_{X_0\supseteq X}\left(\prod [X_0]_F\right)$. 
\end{itemize}
}
The {\bf GCL} then comprises 
$2^{n_F}$ nodes which correspond to $2^{n_F}$ general concepts 
({\bf Proposition~3.5}). 
Specifically, there is no more need to search 
for the object classes that satisfy the requirements for general concepts.
The general extents are nothing but the object classes equipped with non-trivial contextual equivalent classes of attributes.
The {\bf GCL}
thus provides a categorisation for whatever object sets
that can be explicitly labelled by attributes.
Subsequently, the following conjugate relation considerably reduces the complexity for deducing {\bf GCL}.
\prp{
$(X,\rho(X),\eta(X))$ is a general concept if only if  $(X^c, \rho(X^c),\eta(X^c))$ is a general concept,\\
where
$\rho(X^c)=\neg \eta (X)$ and $\eta(X^c)=\neg \rho(X)$.}
\prf{
Since $E_F$ is a sigma algebra by {\bf Proposition~3.5}, $X\in E_F\iff X^c \in E_F$.
Moreover, 
with {\bf Proposition~3.6},\\
$\neg{\rho}(X)=\neg \sum [X]_F= \prod \left\lbrace \mu\mid\ \neg\mu\in [X]_F\right\rbrace=\prod \left\lbrace \mu\mid\ \mu\in [X^c]_F\right\rbrace =\prod [X^c]_F=\eta(X^c)$. Likewise,\\
$\neg{\eta}(X)=\neg \prod [X]_F=\sum [X^c]_F=\rho(X^c)$.}
Interestingly, although the above result  
has a similar appearance to
Eq.~(\ref{eq:rsl_duality}), it possesses a completely 
different meaning.  
The {\bf GCL} is {\it self-dual} in the sense that the general concepts always appear pairwise. 
Subsequently, in order to furnish the lattice structure, the way to order 
concepts as nodes on the lattice remains in question. 
We now proceed to resolve the question.  
\prp{
Given a formal context $F(G,M)$, $\forall X\in E_F$
\begin{itemize}
\item
if $X\not\in  G_{/R}$ and $X\neq\emptyset,\ \rho (X) =\sum_{X_0\subset X } \rho (X_0)$,    
\item
if $X\not\in  G^{/R}$ and $X\neq G,\ \eta (X) =\prod_{X_0\supset X } \eta (X_0)$. 
\end{itemize}
}
\prf{
\begin{itemize}
\item
$\rho(X) =\sum_{X_0\subseteq X}\left(\sum [X_0]_F\right)=\sum [X]_F+\sum_{X_0\subset X}\left(\sum [X_0]_F\right)=\rho(X) +\sum_{X_0\subset X } \rho (X_0)$ by {\bf Proposition 3.6}, hence, $\rho(X)\geq \sum_{X_0\subset X } \rho (X_0)$. In addition, if one can prove that $\sum_{X_0\subset X } \rho (X_0)\geq \rho(X)$ then the statement is established.
Since $X\not\in G_{/R} \cup \left\lbrace \emptyset\right\rbrace$, 
there exists some subset $X_1\subset X$ s.t. 
$X\backslash X_1\neq \emptyset$ and $X_1 \in E_F$.
Consider then $\sum_{X_0\subset X } \rho (X_0)
=\sum_{X_0\subset X } \sum [X_0]_F
=\sum\left(\bigcup_{X_0\subset X} [X_0]_F\right)$. 
Clearly, 
for any {$\tau_1 \in [X_1]_F$}
one has {$\rho(X)\cdot\tau_1\in [X_1]_F\subseteq \bigcup_{X_0\subset X} [X_0]_F$}
because $(\rho(X)\cdot\tau_1)^R=\rho(X)^R\cap \tau_1^R=X_1$. 
Moreover, $(\rho(X)\cdot\neg\tau_1)^R=\rho(X)^R\cap \tau_1^{Rc}=X\backslash X_1$, which implies that $\rho(X)\cdot\neg\tau_1\in [X\backslash X_1]_F\subseteq \bigcup_{X_0\subset X} [X_0]_F$, where notably $[X\backslash X_1]_F\neq \emptyset$ for $[X_1]_F\neq \emptyset$ ({\bf Proposition~3.5}).
Consequently, $\bigcup_{X_0\subset X} [X_0]_F\supseteq \lrbr{\rho(X)\cdot\tau_1,\ \rho(X)\cdot\neg\tau_1}$, which implies that $\sum_{X_0\subset X } \rho (X_0)\equiv\sum\left(\bigcup_{X_0\subset X}[X_0]_F\right)\geq 
(\rho(X)\cdot \tau)+ (\rho(X)\cdot {\neg \tau})\equiv\rho(X)$. 
%
%
\item 
The above result implies that
$\neg \rho (X) =\neg \sum_{X_0\subset X } \rho (X_0)= \prod_{X_0\subset X }\neg \rho (X_0)$ for $X\in E_F \backslash G_{/R}\backslash \left\lbrace \emptyset\right\rbrace$.
Then, by {\bf Lemma 3.7},
$\eta (X^c) = \prod_{X_0\subset X }\eta (X_0^c)$ for $X\in E_F \backslash G_{/R}\backslash \left\lbrace \emptyset\right\rbrace$
which is equivalent to $\eta (X) = \prod_{X_0\supset X }\eta (X_0)$ for $X\in E_F \backslash G^{/R}\backslash \left\lbrace G\right\rbrace$. 
\end{itemize}
}
All the general concepts, i.e. 
$(X,\rho(X),\eta(X))\ \forall X\in E_F$,
can be {deduced in a definite manner} based on a few fundamental attributes.
\prp{
Subject to the formal context $F(G,M)$ $\forall X\in E_F$
\begin{itemize}
\item
if $X\neq \emptyset$ 
then $\rho (X)=\sum_{k\in K}\rho(D_k)$ for some $K\subseteq \left\lbrace 1,\ldots n_F \right\rbrace$,
\item
if $X\neq G$
then $\eta (X)=\prod_{{k}\not\in K}\eta(D_k^c)$ for some $K\subseteq \left\lbrace 1,\ldots n_F \right\rbrace$, where $D_k^c\equiv G\backslash D_k$.
\end{itemize}
}
\prf{
Since $X\in E_F$, {\bf Proposition~3.5} entails that
$X=\bigcup_{k\in K}D_k =\bigcap_{{k}\not\in K}D_k^c$ for some $K\subseteq \left\lbrace 1,\ldots n_F \right\rbrace$.
\begin{itemize}
\item
If $X\not\in G_{/R}$ then $\rho (X)=
\sum_{X_0\subset \bigcup_{k\in K}D_k} \rho (X_0)=\sum_{k\in K}\rho(D_k)$ ({\bf Definition~2.4, Lemma~3.8}).
Meanwhile, {$X\in G_{/R}$} entails that $X=D_k$ and $\rho (X)\equiv \rho(D_k)$ for some $k\in \left\lbrace 1,\ldots n_F \right\rbrace$. 
\item
If $X\not\in G^{/R}$ then
$\eta (X) =\prod_{X_0\supset \bigcap_{{k}\not\in K}D_k^c} \eta (X_0)=\prod_{{k}\not\in K}\eta(D_k^c)$.
Meanwhile, {$X\in G^{/R}$} entails that $X=D_k^c$ and $\eta (X)\equiv \eta(D_k^c)$ for some $k\in \left\lbrace 1,\ldots n_F \right\rbrace$. 
\end{itemize}
}
Thus, once
${\cal P}_\rho:=\left\lbrace \rho(\emptyset)\right\rbrace\cup \left\lbrace \rho(D_k)\mid 1\leq k\leq n_F\right\rbrace$ or 
${\cal P}_\eta:=\left\lbrace \eta(G)\right\rbrace\cup \left\lbrace 
\eta(D_k^c)\mid 1\leq k\leq n_F\right\rbrace$ is obtained, {\bf Proposition~3.9} in practice 
ensures that the full {\it Grsp}'s and {\it Gfcp}'s can be unambiguously determined.
However, the other instructive issue 
is that certain particular attributes can be 
determined without ${\cal P}_\rho$ or ${\cal P}_\eta$. 
\prp{
Given a formal context $F(G,M)$, $\eta(D_k)$ is an $M^\ast$-atom and $\rho(D_k^c)$ an $M^\ast$-coatom ({\bf Lemma~2.12}) for $1\leq k\leq n_F$.
It can be identified that
$\eta(D_k)=\prod \Psi^k$ and $\rho(D_k^c)=\sum ({\breve M}\backslash \Psi^k)$
with $\Psi_M=\lbrace m\in M \mid m\in D_k^I\rbrace\cup
\lbrace \neg m \mid m\not\in D_k^I, m\in M\rbrace$, where
$\Psi^k\in {\bf \Psi}_M$ ({\bf Definition~2.6}).}
\prf{
Since $g^R=D_k^I\ \forall g\in D_k$, 
if $\Psi^k=\lbrace m\in M \mid m\in D_k^I\rbrace\cup
\lbrace \neg m \mid m\not\in D_k^I, m\in M\rbrace$ then
$(\prod \Psi^k)^R=D_k$.
It turns out that $(\prod \Psi^k)\in [D_k]_F$,
and $\eta(D_k)=\prod [D_k]_F\leq \prod \Psi^k$.
Meanwhile, $\prod \Psi^k\succ_M {\bf 0}$ by {\bf Lemma~2.12}, telling that
$\forall \mu \in M^\ast\ \mu\not< \prod \Psi^k$ unless $\mu={\bf 0}$.
Note that $\eta(D_k)\neq {\bf 0}$ for $1\leq k\leq n_F$
since ${\bf 0}^R$ is the empty object set {$\emptyset$}.
Therefore, $\eta(D_k)=\prod \Psi^k$, where $\Psi^k=\lbrace m\in M \mid m\in D_k^I\rbrace\cup \lbrace \neg m \mid m\not\in D_k^I\rbrace$.
On the other hand, $\rho(D_k^c)=\neg \eta(D_k)= \neg\prod \Psi^k =\sum ({\breve M}\backslash \Psi^k)$ by {\bf Proposition~3.7}, where notably ${\breve M}\backslash \Psi^k\in {\bf \Psi}_M$.} 
Notably, {\bf Proposition~3.10} is 
concerned with $\eta(D_k)$ and $\rho(D_k^c)$ 
rather than {$\eta(D_k^c)$ and $\rho(D_k)$} respectively required in 
${\cal P}_\rho$ and ${\cal P}_\eta$.
It is also remarkable that 
based on the results of {\bf Proposition~3.10}
one may further resolve a
truly feasible and realistic construction for {\bf GCL} \cite{LLJD12-2}.
However, 
to recover the RSL- and FCL- nodes in terms of
the {\it general theory of concept lattice} remains the concern for the moment.
Subject to this goal, 
it is more convenient to adopt the irreducible expressions of {\bf Definition~2.14} 
to find the elements of ${\cal P}_\rho$ and ${\cal P}_\eta$, which is described 
in order.
\prp{
Subject to the formal context $F(G,M)$ the {\it Grsp} and {\it Gfcp}
can be expressed in terms of {\it irreducible attribute classes}:
\begin{itemize}
\item
Expressed in terms of {\bf DNF}, the {\it Grsp} can be simplified (Eq.~(\ref{eq:irred_subtract})) as 
\[
\rho(X) =\sum_{X_0\subseteq X}\left(\sum [X_0^{\times}]_F\right)
=\sum_{X_0\subseteq X}\sum \left(\bigcap_{X_0\subset X_i\subseteq X}\frac{[X_0^{\times}]_F}{[X_i^{\times}]_F}\right).
\]
\item
Expressed in terms of {\bf CNF}, the {\it Gfcp} can be simplified as 
\[
\eta (X)=\prod_{X_0\supseteq X}\left(\prod [X_0^{+}]_F\right)
=\prod_{X_0\supseteq X}\prod \left(\bigcap_{X_0\supset X_j\supseteq X}\frac{[X_0^{+}]_F}{[X_j^{+}]_F}\right).
\]
\end{itemize}
}
\prf{
\begin{itemize}
\item
Written in {\bf DNF}, 
$\rho(X)=\sum_{k_\rho} \psi_{k_\rho}$ in which 
$\psi_{k_\rho}$ is a product of members of ${\breve M}$ and 
$\psi_{k_\rho}^R\subseteq X$ since $\rho(X)^R=X$.
Thus, $\rho(X)=\sum_{X_0\subseteq X}\lrp{\sum \varphi_{X_0}}$ 
(cf. {\bf Proposition~3.6}), where 
 $\varphi_{X_0}$ is a subset of $[X_0]_F$
which comprises only the attributes 
given in terms of a product of members of ${\breve M}$.
For the members of $\varphi_{X_0}$, if $\psi_0\not\in_{F\times} M^\ast$ 
then there must be $\psi_1\in \varphi_{X_0}$ such that $\psi_1>\psi_0$ ($\psi_1+\psi_0=\psi_1$) and $\psi_1\in_{F\times} M^\ast$ ({\bf Definition~2.14}). 
Therefore, $\sum \varphi_{X_0}=\sum [X_0^{\times}]_F$, hence, $\rho(X) =\sum_{X_0\subseteq X}\left(\sum [X_0^{\times}]_F\right)$.\\
Moreover, consider a part of expression of 
$\sum_{X_0\subseteq X}\left(\sum [X_0^{\times}]_F\right)$ which emerges as 
$\sum [X_1^{\times}]_F)+\sum [X_2^{\times}]_F$ 
with $X_1\subset X_2$. For $\mu\in [X_1^{\times}]_F$, if 
$\mu=\alpha\cdot \nu$ with $\alpha\in [X_2^{\times}]_F$ or $\nu\in [X_2^{\times}]_F$ 
then $\mu$ is absent in 
$\left(\sum [X_1^{\times}]_F\right)+\left(\sum [X_2^{\times}]_F\right)$
since $\mu+\left(\sum [X_2^{\times}]_F\right)=\left(\sum [X_2^{\times}]_F\right)$.
Consequently, $\sum_{X_0\subseteq X}\left(\sum [X_0^{\times}]_F\right)
=\sum_{X_0\subseteq X}\sum \left(\bigcap_{X_0\subset X_i\subseteq X}\frac{[X_0^{\times}]_F}{[X_i^{\times}]_F}\right)$.
\item
By {\bf Proposition~3.7}, $\eta(X)=\neg \rho(X^c)=\neg \sum [X^c]_F
=\prod \left(\bigcup_{X_0^c\subseteq X^c} \left\lbrace \mu\mid\ \neg\mu=[\left(X_0^c\right)^{\times}]_F\right\rbrace\right)$.
Subsequently, $\neg\mu=[\left(X_0^c\right)^{\times}]_F$ iff $\mu=[\left(X_0\right)^{+}]_F$ by {\bf Lemma~2.15}.
Therefore, $\eta(X)=\prod \left(\bigcup_{X_0\supseteq X} \left\lbrace \mu\mid\ \mu=[\left(X_0\right)^{+}]_F\right\rbrace\right)
=\prod_{X_0\supseteq X}\left(\prod [X_0^{+}]_F\right)$.\\
Similar to the case for $\rho(X)$, consider 
$\prod [X_1^{+}]_F\cdot \prod [X_2^{+}]_F$
in which $X_1\supset X_2$
within the expression $\prod_{X_0\supseteq X}\left(\prod [X_0^{+}]_F\right)$.
For $\mu^\prime\in [X_1^{+}]_F$, if 
$\mu^\prime=\alpha^\prime+\nu^\prime$ with 
$\alpha^\prime\in [X_2^{+}]_F$ or $\nu^\prime\in [X_2^{+}]_F$ then
$\mu^\prime\cdot \left(\prod [X_2^{+}]_F\right)=\left(\prod [X_2^{+}]_F\right)$.
Consequently, $\prod_{X_0\supseteq X}\left(\prod [X_0^{+}]_F\right)
=\prod_{X_0\supseteq X}\prod \left(\bigcap_{X_0\supset X_j\supseteq X}\frac{[X_0^{+}]_F}{[X_j^{+}]_F}\right)$.
\end{itemize}
}
Apparently, the above results are consequences of {\bf Proposition~3.6}.
\prp{
Given a formal context $F(G,M)$, one can
obtain the {\it contextual} {\it truth} and {\it falsity}  
$(1_\eta$ and $0_\rho)$, in contrast to 
the conventional {\it truth} and {\it falsity} $({\bf 1}$ and ${\bf 0})$ in the following way,
\begin{itemize}
\item
 $\eta(\emptyset)={\bf 0}$ and $\rho(G)={\bf 1}$,
\item
$\rho(\emptyset)+\rho(X)=\rho(X)$ and $\rho(\emptyset)\cdot\rho(X)=\rho(\emptyset)$,\\ 
$\eta(G)\cdot \eta(X)=\eta(X)$ and $\eta(G)+ \eta(X)=\eta(G)$, $\forall X \in E_F$. 
\end{itemize}
Accordingly, $0_\rho:=\rho(\emptyset)$ since it behaves like the {\it falsity} for {\it Grsp}, while
$1_\eta:=\eta(G)$ since it behaves like the {\it truth} for {\it Gfcp}. 
In addition, from {\bf Proposition~3.7}, it follows that $1_\eta=\neg 0_\rho$.
Moreover, for the {\it truth} of {\it Grsp} $1_\rho\equiv{\bf 1}$, while for the {\it falsity} of {\it Gfcp} $0_\eta\equiv{\bf 0}$.}
\prf{
\begin{itemize}
\item
$\eta(\emptyset)=\prod_{X\neq \emptyset} \eta\lrp{X} \leq\prod_{k=1}^{n_F}\eta(D_k)={\bf 0}$ by {\bf Proposition~3.8} and {\bf 3.10} since $\eta(D_k)\eta(D_{k^\prime})
={\bf 0}$ for $k\neq {k^\prime}$.\\ 
On the other hand,
$\rho(G)=\neg {\eta}(\emptyset)=\neg {\bf 0}={\bf 1}$ by {\bf Proposition~3.7}. 
\item 
On employing {\bf Proposition 3.6},\\
$\rho(\emptyset)+\rho(X) =\sum [\emptyset]_F+\sum_{X_0\subseteq X}\left(\sum [X_0]_F\right)=\sum_{X_0\subseteq X}\left(\sum [X_0]_F\right)=\rho(X)$,\\
$\rho(\emptyset)\cdot\rho(X) =\sum [\emptyset]_F\cdot\sum_{X_0\subseteq X}\left(\sum [X_0]_F\right)=\sum [\emptyset]_F=\rho(\emptyset)$,\\
$\eta(G)\cdot\eta (X)=\prod[G]_F\cdot\prod_{X_0\supseteq X}\left(\prod [X_0]_F\right)=\prod_{X_0\supseteq X}\left(\prod [X_0]_F\right)=\eta (X)$,\\
$\eta(G)+\eta (X)=\prod[G]_F+\prod_{X_0\supseteq X}\left(\prod [X_0]_F\right)=\prod[G]_F=\eta (X)$.
\end{itemize}}
\crl{
\begin{itemize}
\item
$\rho(\emptyset)\equiv 0_\rho=\sum [\emptyset^{\times}]_F$ and $\eta(G)\equiv 1_\eta=\prod[G^{+}]_F$.
\item
$\rho(D_k)=\rho_0(D_k)
+0_\rho$ and $\eta(D_k^c)=\eta_0(D_k^c)\cdot 1_\eta$,\\ 
where $\rho_0(D_k)=\sum [D_k^{\times}]_F$ and $\eta_0(D_k^c)=
\prod [(D_k^c)^{+}]_F$ for $1\leq k\leq n_F$.
\end{itemize}
}
\prf{
These are direct consequences of {\bf Proposition~3.12},
upon using the formulae in {\bf Proposition~3.11}.
\begin{itemize}
\item 
$\rho(\emptyset)= \sum [\emptyset]_F=\sum [\emptyset^{\times}]_F$ and
$\eta(G)=\prod[G]_F=\prod[G^{+}]_F$. 
\item 
$\rho(D_k)=\sum [D_k^{\times}]_F+ \sum [\emptyset^{\times}]_F=\rho_0(D_k)+0_\rho$ 
and $\eta(D_k^c)=\prod [\left(D_k^c\right)^{+}]_F\cdot \prod[G^{+}]_F=\eta_0(D_k^c)\cdot 1_\eta$. 
\end{itemize}
}
Hence, one may also replace ${\cal P}_\rho$ and ${\cal P}_\eta$
by ${\cal P}^0_\rho:=\left\lbrace 0_\rho\right\rbrace\cup \left\lbrace \rho_0(D_k)\mid 1\leq k\leq n_F\right\rbrace$ and
${\cal P}^0_\eta:=\left\lbrace 1_\eta\right\rbrace\cup \left\lbrace \eta_0(D_k^c)\mid 1\leq k\leq n_F\right\rbrace$, respectively.
Here, one may recover 
the full general concepts
via {\bf Proposition~3.8}
in terms of the {\it irreducible attribute classes} given in {\bf Definition~2.14}.
According to {\bf Proposition~3.9} and based on ${\cal P}^0_\rho$
 and ${\cal P}_\eta$, one then obtains {\it Grsp}'s in {\bf DNF} and 
the {\it Gfcp}'s in {\bf CNF}. We now proceed to consider the corresponding lattice structure. 

The ordering of general concepts can be constructed in an unambiguous way, thereby forming the desired Galois connection.
\prp{
\begin{itemize}
\item
For $X_i,X_j\in E_F,\ X_i\neq X_j$ iff $\rho (X_i)\neq\rho (X_j)$ and $\eta (X_i)\neq\eta (X_j)$.
\item
For $X_i,X_j\in E_F,\ X_i\subset X_j$ iff $\rho (X_i)<\rho (X_j)$ and $\eta (X_i)<\eta (X_j)$.
\end{itemize}
}
\prf{
\begin{itemize}
\item
By {\bf Lemma 2.10}, $\forall \mu_i\forall\mu_j\ \mu_i^R\neq\mu_j^R\implies\ \mu_i\neq\mu_j$.
Here, {$X_i\neq X_j$} means  
$\lsmatr{\rho(X_i)^R\neq\rho(X_j)^R\\ \eta(X_i)^R\neq\eta(X_j)^R}$,
which implies 
$\lsmatr{\rho(X_i)\neq\rho(X_j)\\ \eta(X_i)\neq\eta(X_j)}$.
Thus, $X_i\neq X_j\implies \lsmatr{\rho(X_i)\neq\rho(X_j)\\ \eta(X_i)\neq\eta(X_j)}$.
On the other hand, if {$X_i=X_j$} then
$\lsmatr{\rho(X_i)=\rho(X_j)\\
\eta(X_i)=\eta(X_j)}$.
Consequently,
if $\rho (X_i)\neq\rho (X_j)$ or
$\eta (X_i)\neq\eta (X_j)$ then $X_i\neq X_j$, i.e., $\lsmatr{\rho(X_i)\neq\rho(X_j)\implies X_i\neq X_j\\ 
\eta(X_i)\neq\eta(X_j)\implies X_i\neq X_j}$.
Moreover, 
$\rho(X_i)\neq\rho(X_j)\implies X_i\neq X_j\implies \lsmatr{\rho(X_i)\neq\rho(X_j)\\ \eta(X_i)\neq\eta(X_j)}$
and $\eta (X_i)\neq\eta (X_j)\implies X_i\neq X_j\implies \lsmatr{\rho(X_i)\neq\rho(X_j)\\ \eta(X_i)\neq\eta(X_j)}$.
Therefore, $X_i\neq X_j\iff\lsmatr{\rho(X_i)\neq\rho(X_j)\\ \eta(X_i)\neq\eta(X_j)}$.
\item
With {\bf Definition~3.3}, one 
has the ordering rules\\ $X_i\subseteq X_j \iff (X_j)^{I^\ast}\subseteq (X_i)^{I^\ast}$
and $X_i\subseteq X_j  \iff (X_i)^{\Box^\ast}\subseteq (X_j)^{\Box^\ast}$,\\ 
which is the consequence of Eq.~(\ref{eq:derv_order}) extended to $F^\ast(G,M^\ast)$.
Therefore, 
\begin{eqnarray}
X_i\subset X_j\iff (X_j)^{I^\ast}\subset (X_i)^{I^\ast} 
&\iff& 
\eta(X_i)\equiv\prod (X_i)^{I^\ast} < \prod (X_j)^{I^\ast}\equiv
\eta(X_j),\nonumber\\
X_i\subset X_j\iff (X_i)^{\Box^\ast}\subset (X_j)^{\Box^\ast}\ 
&\iff& \rho(X_i)\equiv\sum (X_i)^{\Box^\ast} < \sum (X_j)^{\Box^\ast}
\equiv 
\rho(X_j),\nonumber
\end{eqnarray}
by {\bf Proposition~3.4}.
\end{itemize}
}
\prp{
The conventional FCL and RSL nodes 
can always be recovered from the {\bf GCL}:
\begin{itemize}
\item
The criterion 
for {$X$ being an RSL extent}, which is
$Y_{rsl}^\Diamond=X$ by $Y_{rsl}:=X^\Box$,
can be satisfied with\\ 
$X\in\lrbr{\bigcup_{m\in M_0}m^R \mid M_0\subseteq M}$.\\
The criterion 
for {$X$ being an FCL extent}, which is $Y_{fcl}^I=X$ by $Y_{fcl}:=X^I$,
can be satisfied with\\ 
$X\in \lrbr{\bigcap_{m\in M_0}m^R \mid M_0\subseteq M}$.
\item
Whenever $X\in E_F$ is an FCL and/or RSL extent,\\
the corresponding RSL intent $Y_{rsl}$ 
can be faithfully read out from $\rho(X)$, \\
the corresponding FCL intent $Y_{fcl}$ can be faithfully read out from $\eta(X)$.
\end{itemize}
}
\prf{
\begin{itemize}
\item
The criteria for $X$ being the extent 
can be rearranged by means of Eq.~(\ref{eq:atderiv}).\\
For $Y_{rsl}=X^{\Box}$, it turns out that $\lrp{X^{\Box}}^\Diamond\equiv \bigcup_{m\in X^{\Box}} m^R=X$. Note that $\forall M_0\subseteq M$ if  
$X=\bigcup_{m\in M_0}m^R$ then $X^{\Box}=M_0$,
thus, $\bigcup_{m\in X^{\Box}} m^R=\bigcup_{m\in M_0}m^R\equiv X$.
Therefore, $\lrp{X^{\Box}}^\Diamond=X$ for $X\in\lrbr{\bigcup_{m\in M_0}m^R \mid M_0\subseteq M}$.
On the other hand, for $Y_{fcl}=X^{I}$, one requires that
$\lrp{X^{I}}^I\equiv \bigcap_{m\in X^{I}} m^R=X$.
Likewise, if $X=\bigcap_{m\in M_0}m^R$ then $M_0$, thus, $\bigcap_{m\in X^{I}} m^R=\bigcap_{m\in M_0}m^R\equiv X$.
Consequently, $\lrp{X^{I}}^I=X$ for $X\in \lrbr{\bigcap_{m\in M_0}m^R \mid M_0\subseteq M}$.
\item
By {\bf Proposition~3.11},
\[\rho(X)=\sum_{X_0\subseteq X}
\sum \left(\bigcap_{X_0\subset X_i\subseteq X}\frac{[X_0^{\times}]_F}{[X_i^{\times}]_F}\right)
=\sum
\left(\bigcup_{X_0\subseteq X} \left(\bigcap_{X_0\subset X_i\subseteq X}\frac{
[X_0^{\times}]_F}{[X_i^{\times}]_F}\right)\right), 
\]
where $\bigcup_{X_0\subseteq X} \left(\bigcap_{X_0\subset X_i\subseteq X}\frac{[X_0^{\times}]_F}{[X_i^{\times}]_F}\right) \supset \bigcup_{X_0\subseteq X}\lrbr{m\in M\mid m^R=X_0}$ 
according to {\bf Lemma~2.15}. 
Moreover, 
$\bigcup_{X_0\subseteq X}\lrbr{m\in M\mid m^R=X_0}= \left\lbrace m\in M\mid m^R\subseteq X\right\rbrace
=X^{\Box}$ by Eq.~(\ref{eq:atderiv}). Therefore, 
if $m\in X^{\Box}\equiv Y_{rsl}$ then $m\in \bigcup_{X_0\subseteq X} \left(\bigcap_{X_0\subset X_i\subseteq X}\frac{[X_0^{\times}]_F}{[X_i^{\times}]_F}\right)$.
Similarly, 
\[
\eta (X)=\prod\left(\bigcup_{X_0\supseteq X} \left(\bigcap_{X_0\supset X_j\supseteq X}\frac{[X_0^{+}]_F}{[X_j^{+}]_F}\right)\right),
\]
where $\bigcup_{X_0\supseteq X} \left(\bigcap_{X_0\supset X_j\supseteq X}\frac{[X_0^{+}]_F}{[X_j^{+}]_F}\right)\supset \left\lbrace m\in M\mid m^R\supseteq X\right\rbrace=X^I$ by {\bf Lemma~2.15} and Eq.~(\ref{eq:atderiv}). 
Therefore, if $m\in X^I\equiv Y_{fcl}$ then $m\in \bigcup_{X_0\supseteq X} \left(\bigcap_{X_0\supset X_j\supseteq X}\frac{[X_0^{+}]_F}{[X_j^{+}]_F}\right)$.
\end{itemize}
}
Remarkably, the roles the RSL and FCL play within the 
{\bf GCL} can be understood easily.
The components of an RSL intent are those {\it simple} attributes 
({\bf Definition~2.5}) found from the {\it Grsp} in {\bf DNF},
whereas the components of an FCL intent are {\it simple} attributes found
from {\it Gfcp} in {\bf CNF}.
Moreover, the RSL is only able to categorise those object classes
which are expressible in terms of a {\it union} of $m^R$'s for some $m$'s in $M$,
whereas the FCL is only able to categorise those object classes
which are expressible in terms of an {\it intersection} of $m^R$'s for some $m$'s in $M$. 

One may construct the {\bf GCL} in terms of lattice theory 
\cite{Wi82,YY04,BH00}.
\prp{
Based on the collection of all the general concepts 
{$L_F=\lrbr{(X,\rho(X),\eta(X))\mid\ X\in E_F}$},
the {\bf GCL} 
exhibits a lattice structure $\Gamma_F:=(L_F,\wedge,\vee,\dagger)$, where 
{$\wedge$}~(meet) and {$\vee$}~(join) are
binary operators and the unary operator {$\dagger$} provides 
a kind of {\it duality}: 
\begin{itemize}
\item
$L_F$ defines 
the partially ordered set (poset)
$(L_F,\leq)$ with the unambiguous order {$\leq$}.
\item
The lattice operations {$\wedge$, $\vee$ and $\dagger$}
are well defined.\\
$\Gamma_F$ is self-dual in the sense 
that $l\in L_F$ iff $l^\dagger\in L_F$. 
{$\dagger$} reverses the ordering of lattice nodes 
in the sense that  $l_i\leq l_j$ iff $l_i^\dagger \geq l_j^\dagger$. 
\item
$\Gamma_F$ is complete in the sense that the supremum and infimum are found to be
$l_{sup}=(G,{\bf 1},1_\eta)$ and $l_{inf}=(\emptyset,0_\rho,{\bf 0})$, respectively, where $G$ is the full collection of objects.
\end{itemize}
}
\prf{
For all $l_k \in L_F$, let $l_k=(X_k,\rho(X_k),\eta(X_k))$.
\begin{itemize}
\item
Because
$X_i\subseteq X_j$ iff $\left\lbrace\begin{smallmatrix}
\rho (X_i)\leq\rho (X_j) \\
\eta (X_i)\leq\eta (X_j)
\end{smallmatrix}\right.$ ({\bf Proposition~3.14}),
it is natural to define $l_i\leq l_j$ via 
$\left\lbrace\begin{smallmatrix}
X_i\subseteq X_j\\
\rho (X_i)\leq\rho (X_j) \\
\eta (X_i)\leq\eta (X_j)
\end{smallmatrix}\right.$.
\item
$\forall l_i\forall l_j\forall l_k\in L_F$, one has 
\begin{eqnarray}
l_i\wedge l_j &:=& \left(X_i\cap X_j,\rho(X_i\cap X_j),\eta (X_i\cap X_j)\right)
\in L_F\ \mbox{since}\ X_i\cap X_j\in E_F,\nonumber\\
l_i\vee l_j &:=& \left(X_i\cup X_j,\rho(X_i\cup X_j),\eta (X_i\cup X_j)\right)
\in L_F\ \mbox{since}\ X_i\cup X_j\in E_F,\nonumber\\
l_k^\dagger&:=& (X_k^c, \rho (X_k^c), \eta(X_k^c))\equiv (X_k^c,\neg \eta(X_k),\neg \rho (X_k))\ \mbox{since}\ X_k^c\in E_F,\nonumber
\end{eqnarray}
where $(X_k^c, \rho (X_k^c), \eta(X_k^c))
=(X_k^c,\neg \eta(X_k),\neg \rho (X_k))$ is based on {\bf Proposition~3.7}.\\
Moreover, $l_i\leq l_j$ iff $l_i^\dagger \geq l_j^\dagger$ as
$\left\lbrace\begin{smallmatrix}
X_i\subseteq X_j\\
\rho (X_i)\leq\rho (X_j) \\
\eta (X_i)\leq\eta (X_j)
\end{smallmatrix}\right.\iff 
\left\lbrace\begin{smallmatrix}
X_i^c\supseteq X_j^c\\
\rho (X_i^c)\geq\rho (X_j^c) \\
\eta (X_i^c)\geq\eta (X_j^c)
\end{smallmatrix}\right.$ or 
$\left\lbrace\begin{smallmatrix}
X_i^c\supseteq X_j^c\\
\neg \rho (X_i)\geq\neg \rho (X_j) \\
\neg \eta (X_i)\geq\neg \eta (X_j)
\end{smallmatrix}\right.$. 
\item
Given $L_0\subseteq L_F$, let $l_0=(X_0,\rho(X_0),\eta(X_0))\in L_0$
in which $X_0\in E_0\subseteq E_F$. Then,
\begin{eqnarray}
  \bigvee_{l_0\in L_0} l_0 &=& \left(\bigcup_{X_0\in E_0}X_0,\rho\lrp{\bigcup_{X_0\in E_0}X_0},\eta\lrp{\bigcup_{X_0\in E_0}X_0}
 \right) \in L_F,\nonumber\\
 \bigwedge_{l_0\in L_0} l_0&=& \left(\bigcap_{X_0\in E_0}X_0,
 \rho\lrp{\bigcap_{X_0\in E_0}X_0},\eta\lrp{\bigcap_{X_0\in E_0}X_0}\right) \in L_F.\nonumber
\end{eqnarray}
Since  $\bigvee_{l_0\in L_0} l_0\geq\ l\ \geq \bigwedge_{l_0\in L_0} l_0\quad \forall l \in L_0$, it turns out that 
\begin{eqnarray}
 l_{sup}&=&\bigvee_{l_0\in L_F} l_0=(G,{\bf 1},\eta(G))\equiv (G,{\bf 1},1_\eta)\nonumber\\
l_{inf}&=&\bigwedge_{l_0\in L_F} l_0 =(\emptyset,\rho(\emptyset),{\bf 0})\equiv (\emptyset,0_\rho, {\bf 0})\nonumber
\end{eqnarray}
by {\bf Proposition~3.8,~3.12}. 
\end{itemize}
}
It is also noteworthy that 
within the expression 
{$(X,\rho(X),\eta(X))$ 
one may regard $(\rho(X),\eta(X))$ as a prescription for 
the {\it general intent} corresponding to the {general extent} $X$.
Such a prescription is convenient for the discussion concerning the RSL and FCL in relation to the {\bf GCL}.
\section{discussion}\label{four}

One has proved the existence 
of the general concept lattice above 
the conventional framework of 
RSL and FCL, following the anticipated 
accomplishments {\bf G1-G5} proposed in Sec.~\ref{one}.
The formal context $F(G,M)$ automatically induces the extended 
formal context $F^\ast(G,M^\ast)$ ({\bf Lemma~2.8}).
In effect, based on $F(G,M)$ there 
are intrinsic relations between 
the objects in $G$ and  
the members of the full generalised 
attribute set $M^\ast$, 
which all pertain to the indispensable structure of 
the content information. 
It is by virtue of $F^\ast(G,M^\ast)$ 
that one can assure 
the 2-tuple {\it general concept} expression in terms of  
$(\mbox{general extent}, \mbox{general intent})$
for {\bf G1},
where the general extents emerge as
the object classes allowable within {\bf GCL}
and the general intents their corresponding 
property descriptions.
From {\bf Proposition~3.4} and {\bf 3.6}
one notices that the general intent can be  
represented by the pair  \footnote{ 
This is a fact that is further supported 
by {\bf Proposition~3.14}, which entails that both
the generalised rough set property and 
generalised formal concept property are ordered simultaneously.
}
\begin{center}
({generalised rough set property}, {generalised formal concept property}) 
, i.e. ({\it Grsp},{\it Gfcp}). 
\end{center}
Moreover, 
the general extents are found to be the collection 
$E_F=\sigma (\lbrace m^R\mid m\in M\rbrace)$ 
in view of {\bf Proposition~3.5} and {\bf Proposition~3.6}, where in particular
$X\in E_F$ iff $X^c\in E_F$.
Note that {\bf G3} is achieved as follows.
Since $E_F$ in effect
comprises whatever distinctive object classes, making
the {\bf GCL} constitute a Hasse diagram 
with $2^{|\sigma (G_{/R})|}=2^{n_F}$ nodes ({\bf G2}),
it is certain on the {\bf GCL} that 
one can always find all the nodes corresponding to the object classes categorised according to the RSL and the FCL, respectively.  
Indeed,
{\bf Proposition~3.15} also provides systematic ways to regain 
both the RSL intent
and FCL intent 
from the irreducible expressions of {\it Grsp} and {\it Gfcp}, 
respectively. 
Another point is that the {\bf GCL} 
permits deterministic construction ({\bf G4}), as is the method 
proposed by {\bf Lemma~3.8,~3.11} and {\bf Corollary~3.13}.
Though to complete the general concept construction in such a manner 
turns out to be inefficient, the method is
rather elucidative for the {\bf GCL} in relation to the conventional lattices.
Especially, one may notice that
the freedom involved in the formal context
can be significantly reduced.
The {\bf GCL} in fact emerges as a self-dual lattice
bearing the structure of {\bf Proposition~3.16}
based on the one-to-one correspondence 
$\forall X\in E_F \lsmatr{\rho (X^c)=\neg \eta (X)\\
\eta (X^c)=\neg \rho (X)}$
demonstrated in {\bf Lemma~3.7}, which
is the conjugateness relation needed by {\bf G5}.
Obviously, the self-duality halves the 
complexity for completing the general intents.


%
Being an original construction, 
the present approach of {\bf GCL} has thus far focused 
on its comparabibility with the existing approaches.
To this end, one found 
the employment of {\it Grsp} in {\bf DNF}
and {\it Gfcp} in {\bf CNF} particularly instructive for 
carrying out such comparisons.
\begin{itemize}
\item
$\rho(X)$ and $\eta(X)$
can be regarded respectively as the {\it generalisation} for 
RSL intent and the {\it generalisation} for FCL intent 
({\bf Proposition~3.1,~3.4}):\\ 
Without altering its theoretical content, 
one may replace the RSL intent 
$X^{\Box}$ by its {\it disjunction} $\sum X^{\Box}$
while $\rho(X)=\sum X^{\Box^\ast}$, which 
can be intuitively arranged into an expression in {\bf DNF}.\\
The FCL intent $X^I$ can be represented 
by its {\it conjunction} $\prod X^I$ while
$\eta(X)=\prod X^{I^\ast}$, which 
can be arranged into an expression in
{\bf CNF}.
\item
One has acquired 
more precise descriptions for the properties of the object class 
through $\rho(X)$ and $\eta(X)$
than through the RSL intent and FCL intent:\\ 
Observe that $\rho(X)=\sum X^{\Box^\ast}=\sum X^{\Box}+(\ldots)\geq 
\sum X^{\Box}$ since obviously $X^{\Box^\ast}\supseteq X^{\Box}$.
If $X$ is an RSL extent, 
$\rho(X)$ suggests that there are in fact 
more properties 
peculiar to the class $X$ to be considered,
in contrast to $\sum X^{\Box}$.\\ 
Likewise, $\eta(X)=\prod X^{I^\ast}\cdot (\ldots)\leq \prod X^I$ 
($\because\ X^{I^\ast}\supseteq X^I$).
Namely, if $X$ is an FCL extent then
$\eta(X)$, in contrast to $\prod X^I$, 
in fact incorporates more properties
the objects in the class $X$ can possess in common.
\end{itemize}
Technically, the general concept construction reflecting 
the above conventional wisdom of the traditional FCL and RSL
can be accomplished via {\bf Lemma~3.8,~3.11} and 
{\bf Corollary~3.13}, which requires 
the determination of various 
irreducible attributes ({\bf Definition~2.14}). 
Unfortunately, such a method remains tedious and impractical even
though the irreducible attributes are constructable. 
However, things turn out drastically different when 
one instead considers the {\it Gfcp} 
in {\bf DNF} and the {\it Grsp} in {\bf CNF}. 
In Ref.~\cite{LLJD12-2} it will be shown that
the {\it Gfcp} can be identified 
as components of the {\it contextual truth} $1_\eta$ 
and $\rho(X)$ the {\it contextual falsity} 
$0_\rho$ defined in {\bf Proposition~3.12}.
Moreover, the determination
of the general intents can be implemented by binary representations 
which label the object classes. 
Noteworthy is also that the {contextual truth} (or {falsity}) also plays the crucial role for the logical deduction in the frame of classical {\it logics}. In practice, all the {\it rules of implication} concerned with the attributes will be shown to be derivable based on $\mu\rightarrow \mu\cdot 1_\eta\ \forall \mu$ (or, equivalently, 
from $\mu+0_\rho \rightarrow \mu$).  
We shall present the tractability 
in both obtaining the full general concepts 
and determining all the accompanied rules of logic deduction for the general concept lattice in the next paper \cite{LLJD12-2}. 

In conclusion, the {\bf GCL} provides the comprehensive 
categorisation for whatever distinctive objects according 
to whatever properties one can refer to in the formal context.
The general extents 
are collected as $E_F=\lrbr{\mu^R\mid \mu\in M^\ast}$
as opposed to {\bf Proposition~3.15}, which states 
\footnote{More precisely, 
$E_{F}^{rsl}=\lrbr{\bigcup_{m\in M_0}m^R \mid M_0\subseteq M}\cup\lrbr{\emptyset}$
and $E_{F}^{fcl}=\lrbr{\bigcap_{m\in M_0}m^R \mid M_0\subseteq M}\cup\lrbr{G}$,
as will be further clarified in Ref.~\cite{LLJD12-2}. 
Note that in their constructions
$\emptyset$ (the empty object set) is always included as an RSL extent  
and $G$ as an FCL extent in view of {\it the completness of lattice}
\cite{Wi82,YY04}.}
that
\[
E_{F}^{rsl}=\lrbr{\bigcup_{m\in M_0}m^R \mid M_0\subseteq M},\ 
E_{F}^{fcl}=\lrbr{\bigcap_{m\in M_0}m^R \mid M_0\subseteq M}.
\]
Clearly, 
$E_{F}^{rsl}\cup E_{F}^{fcl}\subset E_F$ 
$E_F$ is also referred to as 
$\lrbr{\bigcup_{D\in E^0} D\mid E^0\subseteq G_{/R}}$ 
({\bf Proposition~3.5}), meaning that the categorisation the {\bf GCL} can accomplish ranges over all 
the subsets of $G$ that 
are distinct from the perspective 
of formal context.  
It is also remarkable that
the {\it general intent}
is more adequately represented by
the equivalent class of attributes  
{$[X]_F=\lrbr{\mu\mid\mu^R=X,\ \mu\in M^\ast}$}
meaning that
$[X]_F$ refers to all the attributes the objects in 
$X$ possess in common.
Here, $\eta(X)$ and $\rho(X)$
simply appear as the bounds of $[X]_F$.
It can be further shown \cite{LLJD12-2} that 
every general intent is {\it distinctive} in the sense 
that $[X_i]_F\cap [X_j]_F=\emptyset$ iff $X_i\neq X_j$ and
$M^\ast=\bigcup_{X\in E_F}[X]_F$.
Thus, the {\bf GCL}
not only provides 
the categorisation for all 
the discernible subsets of $G$ 
but also exhausts 
the full generalised
attribute set $M^\ast$. 

\section*{acknowledgements}
The authors are grateful to 
Dr. Arthur Chen of Tamkang University who asked
for clarifying some mathematical definitions 
from the FCA theory in the year~2012, 
a crucial issue that inspired the present treatise. 
This paper is partially supported by Ministry of Science and Technology, Taiwan 
(Grant number: MOST~105-2633-E-001-001).

\end{document}